\documentclass[acmsmall]{acmart}
\usepackage{algorithmic}
\usepackage{textcomp}
\usepackage{xcolor}
\usepackage{url}
\usepackage{hyperref}

\usepackage{enumitem}
\usepackage{pifont}
\usepackage{subcaption}
\usepackage{graphicx}
\usepackage{ulem}
\usepackage{multirow}
\usepackage{listings}
\usepackage{color}
\definecolor{pblue}{rgb}{0.13,0.13,1}
\definecolor{pgreen}{rgb}{0,0.5,0}
\definecolor{pred}{rgb}{0.9,0,0}
\definecolor{pgrey}{rgb}{0.46,0.45,0.48}
\usepackage{xspace}
\usepackage{tikz}
\usepackage{pgfplots}
\usepackage{tcolorbox}
\usepackage{colortbl}
\usepackage{wasysym}
\pgfplotsset{compat=1.3}
\makeatletter
\DeclareRobustCommand\onedot{\futurelet\@let@token\@onedot}
\def\@onedot{\ifx\@let@token.\else.\null\fi\xspace}
\def\eg{\emph{e.g.,}\xspace} 

\def\ie{\emph{i.e.,}\xspace}

\def\vs{\emph{vs}\onedot}

\def\method{{\sc CodeEditor}\xspace}
\newcommand{\changeline}[1]{\textcolor{black}{#1}}
\newenvironment{change}[0]{\par\color{black}}{\par}
\newcommand{\changeref}[0]{\color{black}}
\AtBeginDocument{%
  \providecommand\BibTeX{{%
    \normalfont B\kern-0.5em{\scshape i\kern-0.25em b}\kern-0.8em\TeX}}}

\setcopyright{acmcopyright}
\copyrightyear{2018}
\acmYear{2018}
\acmDOI{XXXXXXX.XXXXXXX}

\acmJournal{JACM}
\acmVolume{37}
\acmNumber{4}
\acmArticle{111}
\acmMonth{8}

\begin{document}

\title{{\sc CodeEditor}: Learning to Edit Source Code with Pre-trained Models}

\author{Jia Li $\male$}
\email{lijia@stu.pku.edu.cn}
\author{Ge Li}
\authornote{Corresponding author}
\email{lige@pku.edu.cn}
\author{Zhuo Li}
\email{lizhmq@pku.edu.cn}
\author{Zhi Jin}
\email{zhijin@pku.edu.cn}
\authornotemark[1]
\affiliation{
  \institution{Key Lab of High Confidence Software Technology, MoE, School of Computer Science, Peking University}
  \streetaddress{No.5 Yiheyuan Road, Haidian District}
  \city{Beijing}
  \country{China}
}

\author{Xing Hu}
\affiliation{
  \institution{Zhejiang University}
  \streetaddress{No. 1689 Jiangnan Road, Gaoxin District}
  \city{Ningbo}
  \country{China}
}
\email{xinghu@zju.edu.cn}

\author{Kechi Zhang}
\email{zhangkechi@pku.edu.cn}
\author{Zhiyi Fu}
\email{fuzhiyi1129@gmail.com}
\affiliation{
  \institution{Key Lab of High Confidence Software Technology, MoE, School of Computer Science, Peking University}
  \streetaddress{No.5 Yiheyuan Road, Haidian District}
  \city{Beijing}
  \country{China}
}

\renewcommand{\shortauthors}{Li et al.}

\begin{abstract}
Developers often perform repetitive code editing activities (up to 70\%) for various reasons (\eg code refactoring) during software development.
Many deep learning (DL) models have been proposed to automate code editing by learning from the code editing history.
Among DL-based models, pre-trained code editing models have achieved the state-of-the-art (SOTA) results. Pre-trained models are first pre-trained with pre-training tasks and fine-tuned with the code editing task. Existing pre-training tasks mainly are code infilling tasks (\eg masked language modeling), which are derived from the natural language processing field and are not designed for automatic code editing.

In this paper, we propose a novel pre-training task specialized in code editing and present an effective pre-trained code editing model named {\sc CodeEditor}. 
Compared to previous code infilling tasks, our pre-training task further improves the performance and generalization ability of code editing models.
Specifically, we collect lots of real-world code snippets as the ground truth and use a powerful generator to rewrite them into mutated versions. Then, we pre-train our \method to edit mutated versions into the corresponding ground truth, to learn edit patterns.
We conduct experiments on four code editing datasets and evaluate the pre-trained \method in three settings (\ie fine-tuning, few-shot, and zero-shot).
(1) In the fine-tuning setting, we train the pre-trained \method with four datasets and evaluate it on the test data. \method outperforms the SOTA baselines by 15\%, 25.5\%, and 9.4\% and 26.6\% on four datasets.
(2) In the few-shot setting, we train the pre-trained \method with limited data and evaluate it on the test data. \method substantially performs better than all baselines, even outperforming baselines that are fine-tuned with all data.
(3) In the zero-shot setting, we evaluate the pre-trained \method on the test data without training. \method correctly edits 1,113 programs while the SOTA baselines can not work.
The results show that the superiority of our pre-training task and the pre-trained \method is more effective in automatic code editing.
\end{abstract}



\received{20 February 2007}
\received[revised]{12 March 2009}
\received[accepted]{5 June 2009}

\maketitle

\section{Introduction}
\label{sec:introduction}

\begin{figure}[t]
\centering
\includegraphics[width=0.8\linewidth]{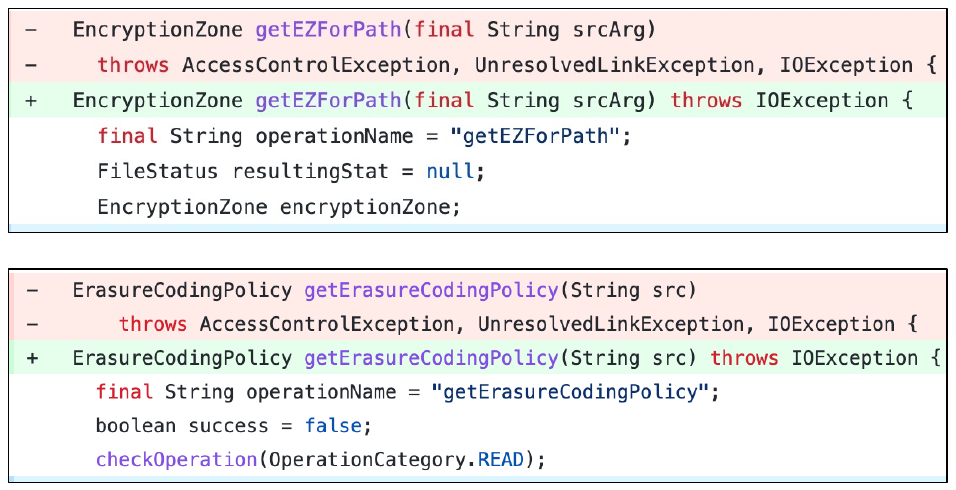}
\caption{Two edits from a real-world software project \cite{CodeChange} in GitHub. They both aim to remove redundant throw exceptions.}
\label{fig:intro_example}
\end{figure}

To improve software systems' stability and maintainability, developers spend lots of effort (\eg more than six hours per week \cite{bosu2013impact}) on editing their source code. For example, developers would modify identifier names or update an outdated API.
A large-scale study in 2,841 Java projects \cite{nguyen2013study} has shown that many edits (up to 70-100\%) follow repetitive patterns.
Figure \ref{fig:intro_example} shows two edits from a real-world software project \cite{CodeChange}. They both aim to remove redundant exceptions (\ie \texttt{AccessControlException and UnresolvedLinkException}) and share an edit pattern.
However, manually designing these repetitive patterns can be tedious and error-prone \cite{ray2013detecting,nguyen2016api}.
Thus, code editing models would be beneficial to save developers' effort by automating code changes learned from previous edit data.

\begin{figure}[t]
\centering
\includegraphics[width=\linewidth]{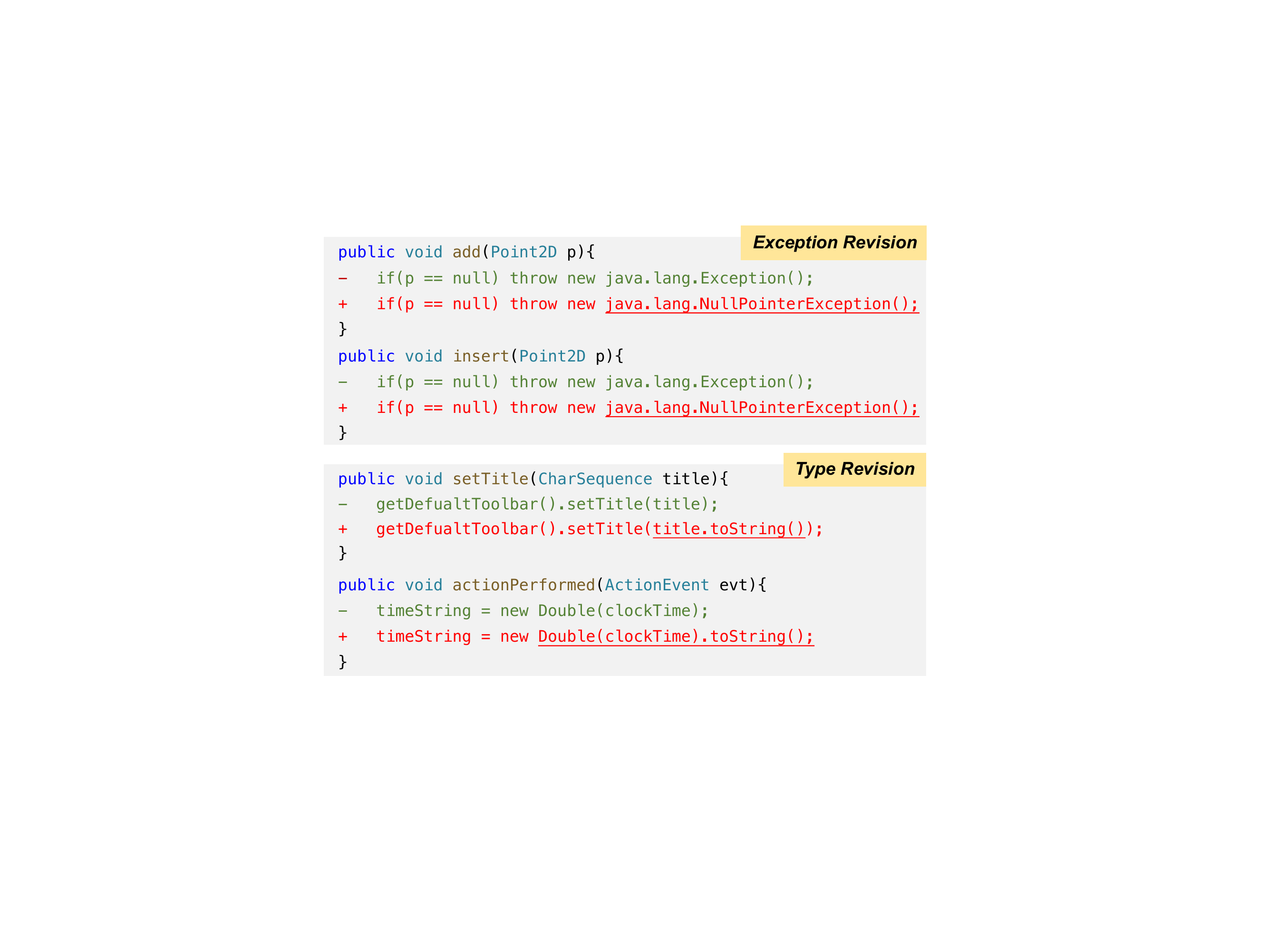}
\caption{\changeline{The comparison of (a) code infilling tasks and (b) our pre-training task. Code infilling tasks randomly mask some tokens or spans in a program and train a model to infill the masked content based on the contexts. Our pre-training task uses a powerful generator to rewrite the code into a mutated version. Then, a model is trained to edit the mutated version into the original version.}}
\label{fig:motivation_ex}
\end{figure}

Recently, deep learning (DL) techniques have been applied to automatic code editing. 
Among DL-based approaches, pre-trained code editing models \cite{tufano2022using,wang2021codet5} have achieved state-of-the-art (SOTA) results on many benchmarks.
Pre-trained models first are pre-trained with \textit{self-supervised pre-training tasks}, and then fine-tuned with the \textit{supervised code editing task}. 
Self-supervision means the labels of training samples are generated automatically without human annotations. Thus, a model can be pre-trained with a large amount of automatically generated data to learn linguistic and commonsense knowledge about the source code.
Nowadays, existing code editing studies \cite{tufano2022using,wang2021codet5} mainly use code infilling tasks (\eg mask language modeling) as the pre-training tasks.
The code infilling tasks randomly mask some tokens or spans in a program and train a model to infill the masked content based on the contexts. 
Figure \ref{fig:motivation_ex} (a) shows a code infilling example. The masked content (\ie \texttt{void}, \texttt{String[]}) is replaced with a specific token (\ie \texttt{MASK}) and highlighted.
Although promising, code infilling tasks are derived from the natural language processing (NLP) field \cite{liu2019roberta,devlin2019bert} and are not designed for automatic code editing. Thus, there are still rooms to improve existing pre-trained code editing models.

In this paper, we propose to extend the conventional code infilling to a novel pre-training task specialized in code editing and present an effective pre-trained code editing model named {\sc CodeEditor}.
Specifically, we first collect lots of programs from open-source communities (\eg GitHub\footnote{https://github.com/}). These programs have passed code reviews and can be viewed as the ground truth.
\changeline{We utilize a powerful \textit{generator} to rewrite these programs into natural but inferior versions (aka mutated versions). 
Then, we pre-train the \method to edit the mutated code into the corresponding ground truth. Figure \ref{fig:motivation_ex} (b) shows a training example in our pre-training task. The rewritten content is highlighted. In Figure \ref{fig:motivation_ex} (b), the generator rewrites a program into a mutated version by modifying two tokens (\texttt{void} $\rightarrow$ \texttt{boolean} and \texttt{String[]} $\rightarrow$ \texttt{int}). Then, the pre-trained model is asked to edit the mutated code into the ground truth.}

Compared to previous code infilling tasks, our pre-training task has two advantages: 
(1) \textbf{Our pre-training task improves the performance of code editing models.} 
The goal of code infilling tasks is to infill a given blank in the source code. Our pre-training task is more challenging and requires a high-level understanding ability to locate inferior parts and a strong generative ability to generate a better alternative. 
Thus, our pre-training task can provide strong supervision signals and further improves the performance of code editing models.
(2) \textbf{Our pre-training task strengthens the generalization ability of code editing models.}
Code infilling tasks are to predict some discrete tokens based on a masked program. Our pre-training task aims to transform a previous program into a new program by automating code changes and is closer to the real-world code editing task.
Thus, our pre-training task endows the model with a practical code editing ability and strengthens the model's generalization ability in real-world code editing.

The key element in implementing our pre-training task is the generator for rewriting programs.
Inspired by previous studies \cite{clark2019electra,zhou2021improving}, we utilize a powerful pre-trained language model for source code - CodeGPT \cite{lu2021codexglue} - as the generator.
CodeGPT is trained on a code corpus consisting of 2.7 million files and is a SOTA language model for source code.
Thus, CodeGPT can derive various informative code snippets that benefit our \method to learn meaningful and diverse edit patterns (\eg API updates, type/object changes, and identifier renaming). These edit patterns resemble the code changes in real-world code editing and thus strengthen the performance of pre-trained models in code editing applications.
\changeline{Specifically, given an original program, we randomly select several spans and replace them with blanks. For each blank, CodeGPT can predict multiple plausible suggestions and ranks them based on probability. 
We consider the original span as the gold and use the first non-gold suggestion to complete the blank.
After completing all blanks, we obtain a mutated version that needs to be edited. Its corresponding original code is the ground truth.}

Following previous studies \cite{chakraborty2021MODIT, tufano2022using}, we conduct experiments on four public code editing datasets.
\changeline{We evaluate our \method in three settings (\ie fine-tuning, few-shot, and zero-shot settings) and employ three widely used metrics, including exact match (EM), CrystalBLEU \cite{eghbali2022crystalbleu}, and edit distance.}
Based on experimental results, we summarize four findings.
(1) In the fine-tuning setting, we train the pre-trained \method with four datasets and evaluate \method on the test data. 
In terms of EM, \method outperforms the SOTA baseline by 15\%, 25.5\%, 9.4\%, and 26.6\% on four datasets.
(2) In the few-shot setting, we sub-sample (\eg 10\%) the datasets, train \method with the limited data, and evaluate \method on the test data. \method still performs substantially better than all baselines, even outperforming baselines that are fine-tuned with all data.
(3) In the zero-shot setting, we evaluate \method on the test data without training. Results show that \method successfully edits 1,123 programs, while baselines generate none of the correct code on four datasets.
(4) We further conduct a case study by analyzing successful cases and failed cases of \method. 

Our main contributions are outlined as follows:
\begin{itemize}
    \item We propose a novel pre-training task for code editing that trains a model to edit an auto-rewritten mutated program into the ground truth. It can improve the performance and generalization ability of code editing models.
    \item We produce a large-scale dataset on our pre-training task and pre-train an effective pre-trained code editing model named {\sc CodeEditor}.
    \item We fine-tune the pre-trained \method with four code editing datasets. Results show that \method is more effective by outperforming the SOTA baseline by up to 26.6\%.
    \item We evaluate the pre-trained \method in zero-shot and few-shot settings. Results show that \method shows a strong generalization ability in both settings.
\end{itemize}

\textbf{Data Availability.}
We open-source our replication package\footnote{https://github.com/LJ2lijia/CodeEditor}, including the datasets and the source code of {\sc CodeEditor}, to
facilitate other researchers and practitioners to repeat our work and verify their studies.

\textbf{Paper Organization.}
Section \ref{sec:background} describes the background of our work. Section \ref{sec:model} presents our model {\sc CodeEditor}. Section \ref{sec:study_design} and Section \ref{sec:result} describe the design and results of our study, respectively. Section \ref{sec:discussion} and Section \ref{sec:related_work} discuss some results and describe the related work, respectively. Section \ref{sec:conclusion} concludes the paper and points out future directions.

\section{Background}
\label{sec:background}

\subsection{Code Editing}
\label{sec:background:code_editing}

Generating the source code using deep learning models has been explored in previous studies \cite{yin2018tranx,sun2020treegen}.
These studies aim to build a model $p(\boldsymbol{y}|\boldsymbol{x})$ that generates the code $\boldsymbol{y}$ based on the given contextual information $\boldsymbol{x}$ (\eg natural language requirements).
In this work, code editing is a special case of $p(\boldsymbol{y}|\boldsymbol{x})$ and contains two main scenarios, including code-to-code editing and comment\&code-to-code editing.

For the code-to-code editing, our goal is to train a model $p(\boldsymbol{y}|\boldsymbol{x})$ that predicts the edited code $\boldsymbol{y}$ (\eg a new version) given the code to be edited $\boldsymbol{x}$ (\eg an old version). For the comment\&code-to-code editing, our target is to train a model $p(\boldsymbol{y}|\boldsymbol{x}, \boldsymbol{z})$ that predicts the edited code $\boldsymbol{y}$ given the code to be edited $\boldsymbol{x}$ and a natural language comment $\boldsymbol{z}$. The comment describes code changes between $\boldsymbol{x}$ and $\boldsymbol{y}$. 

\subsection{Pre-trained Models for Source Code}
\label{sec:background:pre-train}

Recently, pre-trained models for source code \cite{wang2021codet5,niu2022spt-code} have been proposed and achieved SOTA results on many code-related tasks (\eg code generation \cite{SkCoder, AceCoder}, code summarization \cite{EditSum}). Generally, pre-trained models consist of two stages - pre-training and fine-tuning.

During the pre-training, the models are trained with self-supervised pre-training tasks to learn prior knowledge about source code. 
Nowadays, code infilling tasks (\eg mask language modeling, MLM) \cite{feng2020codebert,wang2021codet5} have become the prevalent pre-training tasks for source code.
Code infilling tasks randomly mask some tokens or spans in a program and ask the models to predict the masked content based on the contexts. In this way, code infilling tasks enable the models to learn the syntax of source code.
Then, the pre-trained models are applied to specific tasks (\eg code editing) by fine-tuning with the task-specific data.
Besides fine-tuning, pre-trained models can be applied in few-shot and zero-shot settings. The few-shot setting means that we fine-tune the pre-trained models with a few samples, and the zero-shot setting denotes the pre-trained models are used in specific tasks without training.

The performance of pre-trained models in downstream tasks mainly depends on the knowledge derived from the pre-training tasks.
However, there is limited work exploring pre-training tasks specialized in code editing.
In this paper, we propose a novel pre-training task for code editing and pre-train a code editing model named {\sc CodeEditor}. We evaluate the pre-trained \method in fine-tuning, few-shot, and zero-shot settings.

\subsection{Pre-trained Source Code Language Model}
\label{sec:background:lm}

Language models aim to capture the statistical patterns in languages by assigning occurrence probabilities to a sequence of words. The models will score an utterance high, if it sounds ``natural'' to a native speaker, and score low the unnatural (or wrong) sentences. Programming languages are kinds of languages that contain predictable statistical properties and can be modeled by language models. Given a code token sequence $\boldsymbol{S}=\{s_1, s_2, ..., s_t\}$, the probability of this sequence is computed as:
\begin{equation}
p(\boldsymbol{S})=p\left(s_1\right) p\left(s_2 \mid s_1\right), \ldots, p\left(s_t \mid s_1 s_2, \ldots, s_{t-1}\right)
\end{equation}

Recently, pre-trained language models (PTLMs) have shown to be effective \cite{liu2020multi,lu2021codexglue}. PTLMs are trained with a large-scale corpus of real code files. Given a partial code snippet, they can accurately predict multiple plausible patches. Thus, we utilize a powerful language model to rewrite an original program into a mutated version by replacing some original code spans with natural but inferior alternatives predicted by PTLMs. 
By editing the mutated code into its original version, the pre-trained model can know many diverse and meaningful edit patterns (\eg API updates, identifier renaming).

\section{\method}
\label{sec:model}

\subsection{Overview}
\label{sec:overview}

\begin{figure*}[t]
\centering
\includegraphics[width=\linewidth]{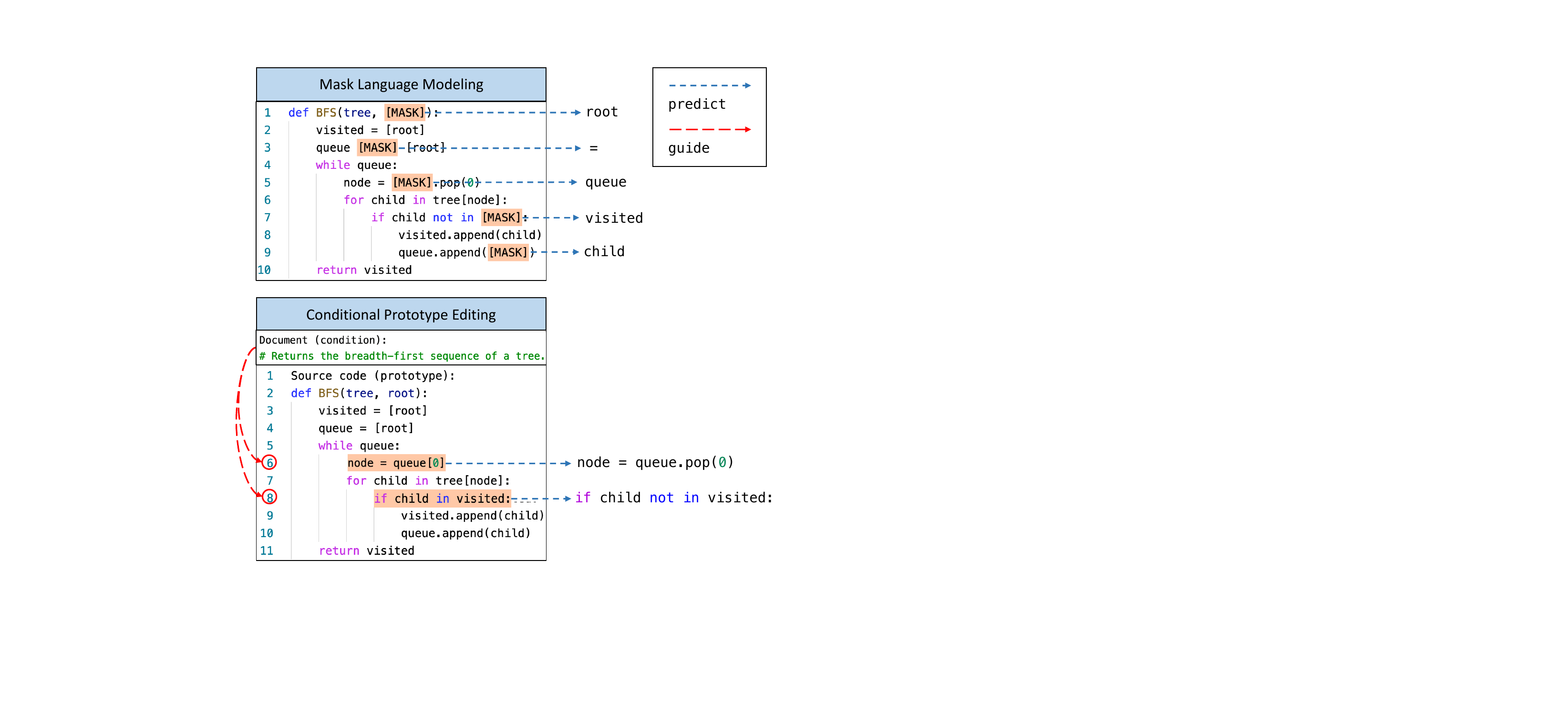}
\caption{The overview of {\sc CodeEditor}.}
\label{fig:overview}
\end{figure*}

The overview of our \method is shown in Figure \ref{fig:overview}.
Our approach comprises three stages: 

\textbf{(1) Producing the pre-training data.}
We first collect lots of real-world programs from open-source communities (\eg GitHub).
\changeline{Then, we use a powerful generator to rewrite these programs into mutated versions.
We view the mutated versions as the code to be edited and the corresponding original versions as the ground truth, to construct the pre-training data.}
The details are explained in Section \ref{sec:model:pseudo_data}.

\textbf{(2) Pre-training the \method.} 
We pre-train our \method with the pre-training data, which takes the code to be edited as inputs and outputs the edited code. 
The details are presented in Section \ref{sec:model:pre-training}.

\textbf{(3) Applying the pre-trained \method.}
Finally, we apply the pre-trained \method to code editing in three settings - fine-tuning, few-shot, and zero-shot. The details are described in Section \ref{sec:model:fine-tune}.

\subsection{Producing the pre-training data}
\label{sec:model:pseudo_data}

The goal of this step is to produce the pre-training data consisting of the code to be edited and the ground truth. 
As shown in Figure \ref{fig:overview}, we first collect lots of individual code snippets from open-source communities (\eg GitHub). These code snippets have been edited carefully and passed strict code reviews. Thus, we consider them as the ground truth.
\changeline{Then, for each code snippet $\boldsymbol{c}$, we utilize a strong generator to rewrite it into a mutated version $\boldsymbol{c'}$.}

Given a code snippet $\boldsymbol{c}=[c_1,c_2,...,c_n]$, $c_i$ denotes $i$-th token and and $n$ is the maximum length.
We randomly select $k$ spans in $\boldsymbol{c}$ and replace them with blanks. This process can be formulated to:
\begin{equation}
\begin{aligned}
    m_{i} \sim \operatorname{unif}\{1, n\} \text { for } i=1 \text { to } k \\
    l_{i} \sim \operatorname{unif}\{a, b\} \text { for } i=1 \text { to } k \\
    s_{i}  = [\boldsymbol{c}_{m_i}, ..., \boldsymbol{c}_{m_i+l_i}] \text { for } i=1 \text { to } k \\
    \boldsymbol{c}^{\text{blank}} = \operatorname{REPLACE}(\boldsymbol{c}, \boldsymbol{s}, \texttt{[BLANK]})
\end{aligned}
\label{equ:span}
\end{equation}
where $\boldsymbol{m}=[m_1,...,m_k]$ and $\boldsymbol{l}=[l_1,...,l_k]$ are the start positions and lengths of $k$ spans, respectively. $\operatorname{unif}\{a, b\}$ denotes a uniform distribution from $a$ to $b$. $\boldsymbol{s}=[s_1,...,s_k]$ is selected $k$ spans.
$\operatorname{REPLACE}(\boldsymbol{c}, \boldsymbol{s}, \texttt{[BLANK]})$ is a function for replacing selected spans $\boldsymbol{s}$ in $\boldsymbol{c}$ with a specific token $\texttt{[BLANK]}$. As shown in Figure \ref{fig:overview} (a), the token \texttt{String[]} is selected and replaced with \texttt{[BLANK]}.

A large-scale study of code edits \cite{nguyen2013study} has found that the sizes of most repetitive edits range from 1 to 6 tokens. Therefore, the $a$ and $b$ in Equation \ref{equ:span} are set to 1 and 6, respectively. 
It means the length of each span is sampled from a uniform distribution of \{1, 6\}.
Previous pre-trained work \cite{zhou2021improving} also has shown that pre-trained models work better when 30\% of tokens are modified. We follow this setting to control the number of spans $k$ so that approximately 30\% of code tokens are selected.

\changeline{Next, we use a generator $p_{G}$ to complete these blanks. For each blank, the generator will output many suggestions that are ranked based on probability. We consider the original span as the gold and pick the first non-gold suggestion as a sub-optimal alternative.}
\begin{equation}
\begin{aligned}
    \hat{s}_{i} \sim p_{G}\left(\boldsymbol{c}^{\text {blank}}\right) \text { for } i=1 \text { to } k \\
    \boldsymbol{c}' = \operatorname{REPLACE}(\boldsymbol{c}, \boldsymbol{s}, \hat{\boldsymbol{s}})
\end{aligned}
\end{equation}
\changeline{where $\hat{\boldsymbol{s}} = [\hat{s}_1,...,\hat{s}_k]$ is selected sub-optimal alternatives predicted by the generator.
Then, we replace original spans $\boldsymbol{s}$ with sub-optimal alternatives $\hat{\boldsymbol{s}}$ to obtain the mutated code $\boldsymbol{c}'$. In Figure \ref{fig:overview} (a), the generator predicts multiple suggestions for the blank, \eg \texttt{String[]}, \texttt{int}. Thus, we select the first non-gold suggestion (\ie \texttt{int}) to complete the blank. Then, the original code snippet is rewritten into a mutated version that has a wrong parameter type.}

\changeline{By repeating the above process, we obtain lots of mutated versions and original versions of code snippets. Then, we consider the mutated versions as the code to be edited $\boldsymbol{x}$ and the corresponding original versions as the ground-truth $\boldsymbol{y}$, to construct our pre-training data.}
If the original code is equipped with a natural language comment $z$, we take the comment as an additional input, building a pre-training sample $(\boldsymbol{x},\boldsymbol{z},\boldsymbol{y})$. Otherwise, the pre-training sample is built as $(\boldsymbol{x},\boldsymbol{y})$.
Our motivation is that these two types of samples correspond to two code editing scenarios. For samples without comments $(\boldsymbol{x},\boldsymbol{y})$, \method learns a code-to-code edit ability that benefits the code-to-code editing scenario. For samples with comments $(\boldsymbol{x},\boldsymbol{z},\boldsymbol{y})$, \method edits the source code with the guidance of comments, which is suitable for the code\&comment-to-code editing scenario.

\subsection{Pre-training the \method}
\label{sec:model:pre-training}

As shown in Figure \ref{fig:overview} (b), this step aims to pre-train \method with the pre-training data produced in Section \ref{sec:model:pseudo_data}.
In this paper, we view the code editing task as a sequence-to-sequence task.
Specifically, \method is trained to generate an output sequence $\boldsymbol{Y}$ based on an input sequence $\boldsymbol{X}$. 
Following previous studies \cite{chakraborty2020codit,tufano2022using}, we build the input and output sequences as follows:
\begin{equation}
\boldsymbol{X} =\left\{\begin{array}{ll}
\boldsymbol{x} & \text {without comments} \\
\boldsymbol{x} \texttt{[SEP]} \boldsymbol{z} & \text {with comments}
\end{array} \quad \boldsymbol{Y}=\boldsymbol{y} \right.
\end{equation}
where \texttt{[SEP]} is a specific token for concatenating two sequences. 

We train \method to generate the output sequence auto-regressively based on the input sequence by minimizing the following loss function:
\begin{equation}
\mathcal{L}(\theta)= - \log P\left(\boldsymbol{Y} \mid \boldsymbol{X} \right) =-\sum_{t=1}^{L} \log P\left(Y_{t} \mid X, Y_{<t}\right)
\label{equ:loss}
\end{equation}
where $Y_{t}$ denotes $t$-th token of the output sequence. 
$\theta$ denotes all trainable parameters and $L$ is the maximum length of the output sequence. 

\subsection{Applying the pre-trained \method}
\label{sec:model:fine-tune}

The pre-trained \method learns a general code editing ability from the pre-training data.
As shown in Figure \ref{fig:overview} (c), the next step is to apply the pre-trained \method to code editing.
In this work, we apply the pre-trained \method in three settings - fine-tuning, few-shot, and zero-shot settings.
In the fine-tuning setting, we train the pre-trained \method with code editing datasets. In the few-shot setting, we sub-sample the datasets and train the pre-trained \method with the limited data. In the zero-shot setting, we directly apply the pre-trained \method to code editing applications without training.

For each setting, we apply the pre-trained \method to two code editing scenarios - code-to-code editing and comment\&code-to-code editing.
We follow the sequence-to-sequence learning and input-output representations used in the pre-training stage. 
For the code-to-code editing scenario, the input sequence is the code to be edited $\boldsymbol{x}$.
For the comment\&code-to-code editing scenario, we concatenate $\boldsymbol{x}$ and a natural language comment $\boldsymbol{z}$ into an input sequence.
In both scenarios, the output sequence is the ground-truth $\boldsymbol{y}$.
Then, we train the pre-trained \method to generate an output sequence based on an input sequence by minimizing the loss function (Equation \ref{equ:loss}).

\subsection{Model Architecture}
\label{sec:model:architecture}

\method is an encoder-decoder neural network based on Transformer \cite{vaswani2017attention}, which contains a bidirectional encoder and an auto-regressive decoder.
Following previous studies \cite{wang2021codet5}, we adopt the same architecture as Text-to-Text-Transfer Transformer (T5) model \cite{raffel2020t5}. (1) The number of layers (\ie Transformer blocks) $L$ = 6. (2) The dimension of the model $d_{model}$ = 512. (3) The dimension of feed-forward layers $d_{ff}$ = 2048. (4) The number of self-attention heads $h$ = 8 and the drop rate $p$=0.1. The total number of parameters is 60M.

\section{Study Design}
\label{sec:study_design}

To evaluate the effectiveness of our {\sc CodeEditor}, we design three research questions (RQ) and perform a large-scale study to answer these questions. In this section, we describe the research questions and details of our study.

\subsection{Research Questions}
\label{sec:study_design:RQ}

We argue that our pre-training task can improve the performance of code editing models.
To prove the effectiveness of our \method, in the RQ1 and RQ2, we fine-tune the pre-trained \method in two code editing scenarios and compare it to SOTA baselines.

\textbf{RQ1: How does \method perform in the code-to-code editing scenario?}

In this RQ, we fine-tune the pre-trained \method with two code-to-code editing datasets, respectively. 
Given a code snippet to be edited, \method is trained to generate a new version by automating generic code changes.
Then, we evaluate the performance of \method on the test data and compare it with SOTA baselines on two datasets.

\textbf{RQ2: How does \method perform in the comment\&code-to-code editing scenario?}

In this RQ, we fine-tune the pre-trained \method with two comment\&code-to-code editing datasets, respectively. 
Given a code snippet to be edited and a natural language comment, the model outputs a new version by automating code changes described by the comment.
We also compare \method with SOTA baselines on the test data.

We hypothesize that our pre-training task endows the model with a practical code editing ability and strengthens the model's generalization ability.
To prove this point, we design the RQ3 that evaluates the pre-trained \method in few-shot and zero-shot settings where the fine-tuning data is limited.

\textbf{RQ3: How does \method perform in few-shot and zero-shot settings?}

For the zero-shot setting, we evaluate the pre-trained \method on the test data without training. For the few-shot setting, we sub-sample (\ie 10\%, 20\%, 30\%, 40\%, 50\%) code editing datasets. We fine-tune the pre-trained \method with the limited data and measure its performance on the test data.

\begin{table}[t]
\caption{Pre-training and fine-tuning datasets (\# instances)}
\begin{tabular}{lccc}
\toprule
\textbf{Dataset}       & \textbf{train}   & \textbf{dev} & \textbf{test}  \\ \midrule
\textbf{Pre-training}  &         &       &       \\
\quad \textit{CodeSearchNet \cite{husain2019codesearchnet}} & 1,519,885 & 50,000 &  -     \\
\textbf{Code-to-code editing}   &         &       &       \\
\quad \textit{Small \cite{tufano2019learning}}         & 46,628   & 5,828  & 5,831  \\
\quad \textit{Medium \cite{tufano2019learning}}        & 52,324   & 6,542  & 6,538  \\
\textbf{Comment\&code-to-code editing}   &         &       &       \\
\quad \textit{Tufano et al. \cite{tufan2021towards}} & 13,670   & 1,713  & 1,704  \\
\quad \textit{new\_large \cite{tufano2022using}}    & 134,209  & 16,773 & 16,780 \\
\bottomrule
\end{tabular}
\label{tab:datasets}
\end{table}

\subsection{Datasets}
\label{sec:study_design:dataset}

Our \method involves two steps - pre-training and fine-tuning. For the pre-training, we need a large-scale corpus consisting of real-world code snippets for producing the pre-training data. For the fine-tuning, we require the code editing datasets containing real-world edit pairs for transferring the pre-trained {\sc CodeEditor}.

\subsubsection{Pre-training}
Following previous pre-trained studies \cite{feng2020codebert,guo2020graphcodebert,wang2021codet5,niu2022spt-code}, we select the CodeSearchNet-java \cite{husain2019codesearchnet}, which is collected from the popular open-source repositories from GitHub. The CodeSearchNet-java dataset contains 1.5 million individual Java methods, including 499,618 methods with comments and 1,070,267 methods without comments.
\changeline{For each method, we follow the pipeline in Section 3.2 and obtain a mutated version. Then, we consider the modified methods as an input and the original methods as an output, to construct a pre-training sample.}

We split the pre-training data into a train set and a dev set. The data statistics are shown in Table \ref{tab:datasets}. To ensure a fair comparison with previous studies, we do not use additional data for pre-training. 

\subsubsection{Fine-tuning}
In our experiments, we apply the pre-trained \method to two code editing scenarios, \ie code-to-code editing, and comment\&code-to-code editing. For each scenario, we select two public datasets for fine-tuning.

\textbf{Code-to-code editing.} We use two Java code editing datasets proposed by Tufano et al. \cite{tufano2019learning}. Tufano et al. extracted method-level pairs from commits in GitHub repositories. Each pair is composed of a previous version and a new version of a Java method.
Based on the length of extracted methods, Tufano et al. provided two datasets: Small and Medium datasets, with the former having a code length below 50 tokens and the latter having a code length between 50-100 tokens. 
\changeline{Note that we use the raw version of two datasets.}
The data statistics are shown in Table \ref{tab:datasets}.

\textbf{Comment\&code-to-code editing.} We use two Java code editing datasets proposed by Tufano et al. \cite{tufan2021towards,tufano2022using}. The datasets are collected from Java open-source projects on GitHub. Tufano et al. further 
filtered low-quality projects and extracted triplets as samples.
Each triplet contains a previous version and a new version of a Java method, and a natural language comment that describes code changes. 
The data statistics are shown in Table \ref{tab:datasets}.

\subsection{Evaluation Metrics}
\label{sec:study_design_metric}

\begin{change}
Following previous code editing studies \cite{thongtanunam2022autotransform, raffel2020t5, chakraborty2021MODIT}, we use the \textit{exact match}, \textit{CrystalBLEU}, and \textit{Edit distance} as evaluation metrics. We regard the code edited by models as a prediction and the code edited by human developers as the ground truth.
\begin{itemize}
    \item \textbf{Exact match (EM)} is the percentage of predictions that has the same token sequence as the ground truth.
    \item \textbf{CrystalBLEU} \cite{eghbali2022crystalbleu} is a metric to precisely and efficiently evaluate code similarity despite trivially shared n-grams. CrystalBLEU is an extension of BLEU \cite{papineni2002bleu} that removes trivially shared n-grams before computing the n-gram overlap between two pieces of code. Extensive experiments show that CrystalBLEU provides higher distinguishability than standard BLEU on the code.
    We reuse official implementations of CrystalBLEU and use training sets of experimental datasets to extract trivially shared n-grams.
    \item \textbf{Edit distance} is the minimum number of token edits (insertions, deletions, or substitutions) needed to convert a prediction into the ground truth. The lower the Levenshtein distance, the closer the prediction is to the ground truth. 
\end{itemize}
\end{change}
    
\subsection{Baselines}
\label{sec:study_design_baseline}
We select 13 recently proposed code editing models as baselines. They can be divided into two categories: baselines trained from scratch and pre-trained baselines. 

\textbf{Baselines trained from scratch} denote that baselines are randomly initialized and further trained with code editing datasets.
\begin{itemize}
    \item \textbf{Tufano et al. \cite{tufano2019learning}} is a pioneer work that utilizes a DL-based model to learn code changes from pull requests.
    \item \textbf{Tufano et al. \cite{tufan2021towards}} proposes two DL-based code editing models that can automate generic code changes and changes recommended by reviewers' comments.
    \item \textbf{CODIT \cite{chakraborty2020codit}} is a tree-based code editing model that leverages the rich syntactic structure of code and generates syntactically correct changes.
    \item \textbf{Transformer \cite{vaswani2017attention}} is a popular DL-based model and has obtained promising results in code-related tasks.
    \item \textbf{AutoTransform \cite{thongtanunam2022autotransform}} is a variant of Transformer. It introduces a Byte-Pair Encoding (BPE) algorithm \cite{sennrich2016neural} to handle unseen tokens and utilizes the self-attention mechanism to model long sequences.
\end{itemize}

\textbf{Pre-trained baselines} are firstly pre-trained with several pre-training tasks and then fine-tuned with the code editing task. Nowadays, pre-trained code editing models have achieved SOTA results on many benchmarks.
\begin{itemize}
    \item \textbf{RoBERT (code) \cite{liu2019roberta}} is a pioneer pre-trained model for natural languages. We continually pre-train it with the source code for comparison.
    \item \textbf{CodeBERT \cite{feng2020codebert}} is a classic pre-trained model for source code. It applies the pre-training tasks for natural languages to the source code and has been widely used in code generation.
    \item \textbf{GraphCodeBERT \cite{guo2020graphcodebert}} is a augmented variant of CodeBERT. It proposes two new pre-training tasks to learn the relationships between the data flow graph and code tokens.
    \item \textbf{CodeGPT \cite{lu2021codexglue}} is a variant of GPT-2 \cite{radford2019gpt-2} that is a powerful pre-trained model for natural language generation. CodeGPT is initialized with GPT-2 and continually pre-trained with the source code.
    \item \textbf{SPT-Code \cite{niu2022spt-code}} is a sequence-to-sequence pre-trained model for source code. It utilizes three pre-training tasks to learn the lexical and syntactic knowledge of source code.
    \item \textbf{T5-review \cite{tufano2022using}} applies a popular pre-trained model - T5 \cite{raffel2020t5} for natural languages into source code and has obtained promising results in code editing.
    \item \textbf{MODIT \cite{chakraborty2021MODIT}} is designed for the comment\&code-to-code editing. MODIT considers commit messages as intents for guiding code editing models.
    \item \textbf{CodeT5 \cite{wang2021codet5}} is a recently proposed pre-trained model for source code. CodeT5 employs identifier-aware pre-training tasks and has achieved SOTA results on many code-related tasks.
\end{itemize}

Note that some baselines (\ie Tufano et al. \cite{tufano2019learning}, CODIT, and AutoTransform) are designed for the code-to-code editing scenario, and MODIT only works in the comment\&code-to-code editing scenario. Thus, we compare our \method to these special baselines in specific scenarios.
Besides, considering that some pre-trained baselines (\ie RoBERTa, CodeBERT, and GraphCodeBERT) only contain an encoder, we follow previous work \cite{chakraborty2021MODIT} and add a six-layer transformer decoder along with these baselines to support code editing. The decoder is randomly initialized and optimized during fine-tuning.

\subsection{Implementation Details}
\label{sec:study_design:implementation}

\subsubsection{Pre-training Details}
We implement our \method with a deep learning development framework - Pytorch\footnote{https://pytorch.org/} and a python package - transformers\footnote{https://huggingface.co/docs/transformers/index}.
We use Byte-Pair Encoding (BPE) \cite{sennrich2016neural} to tokenize the source code and natural language comments into tokens.
The network architecture of \method is described in Section \ref{sec:model:architecture}.
We initialize our model with the pre-trained CodeT5-small \cite{wang2021codet5} (60M), and continually pre-train the model with our pre-training task. 
Note that initializing with existing pre-trained weights is common in previous studies \cite{guo2020graphcodebert,lu2021codexglue,chakraborty2021MODIT}. To make a fair comparison, we also reuse the pre-trained CodeT5-small. This also reduces recent concerns \cite{strubell2019energy} about the carbon footprint and energy consumption caused by the pre-training from scratch. 
We pre-train our \method on a cluster of 32 NVIDIA A100 GPUs with 40G memory. We set the maximum input and output sequence lengths to 512. We pre-train the model for 180,000 steps, with a batch size of 512, and a learning rate of 5e-5 with a linear warm-up for the first 10,000 steps. 
\changeline{Every 10,000 pre-training steps, we measure the performance of our model by calculating the loss of our model on the dev set, but without updating the parameters. Finally, we select the checkpoint with a minimum loss on the dev set as the pre-trained \method.}

\subsubsection{Fine-tuning Details}
During the fine-tuning step, we set the batch size to 32 and the learning rate to 5e-5. The warmup steps are 1000 steps. During testing, we use the beam search and set the beam size to 10.

\section{Results and Analyses}
\label{sec:result}

In this section, we answer three research questions (Section \ref{sec:study_design:RQ}) based on our experimental results.

\subsection{Performance in the code-to-code editing scenario}
\label{sec:result:RQ1}

\textbf{RQ1: How does \method perform in the code-to-code editing scenario?}

\textbf{Motivation.} We investigate whether our \method learns a strong code editing ability to support the code-to-code editing application.

\textbf{Setup.} We fine-tune or train baselines and the pre-trained \method with two code-to-code editing datasets (\ie Small and Medium) described in Section \ref{sec:study_design:dataset}. Then, we use metrics presented in Section \ref{sec:study_design_metric} to evaluate the performance of baselines and our \method on the test data.
Since MODIT \cite{chakraborty2021MODIT} only works in the comment\&code-to-code editing scenario, we omit it in this experiment.

\begin{table*}[t]
\caption{\changeline{The performance of baselines and our \method on two code-to-code editing datasets.} The values in parentheses are relative improvements compared to the SOTA baseline.}
\changeref{}
\resizebox{\linewidth}{!}{
\begin{tabular}{cl|ccc|ccc}
\toprule
\multirow{2}{*}{Type} &
  \multirow{2}{*}{Approaches} &
  \multicolumn{3}{c|}{Small dataset} &
  \multicolumn{3}{|c}{Medium dataset} \\ 
  & & Exact match $\uparrow$ & CrystalBLEU $\uparrow$ & Edit distance $\downarrow$
  & Exact match $\uparrow$ & CrystalBLEU $\uparrow$ & Edit distance $\downarrow$  \\ \midrule
\multirow{5}{*}{\begin{tabular}[c]{@{}c@{}}Trained from\\ Scratch\end{tabular}} 
& Tufano et al. \cite{tufano2019learning}
& 3.45 & 63.36 & 31.11 & 1.62 & 76.23 & 28.53 \\
& Tufano et al. \cite{tufan2021towards}
& 3.91 & 63.95 & 30.86 & 1.84 & 76.71 & 27.79 \\
& CODIT 
& 6.53 & 65.82 & 22.40 & 3.27 & 75.58 & 25.75 \\
& Transformer 
& 7.01 & 68.40 & 25.21 & 4.19 & 78.73 & 27.65 \\
& AutoTransform
& 7.65 & 70.24 & 24.20 & 4.51 & 79.93 & 25.97 \\
\midrule
\multirow{8}{*}{\begin{tabular}[c]{@{}c@{}}Pre-trained \\ Model \end{tabular}}
& RoBERTa (code) & 13.75 & 76.48 & 17.50 & 4.67 & 82.62 & 23.45 \\
& CodeBERT & 14.70 & 77.21 & 15.18 & 4.97 & 84.01 & 22.64 \\
& GraphCodeBERT & 15.91 & 78.34 & 15.30 & 7.49 & 84.49 & 22.48 \\
& CodeGPT & 15.09 & 77.69 & 15.08 & 5.29 & 83.07 & 26.07 \\
& T5-review & 14.82  & 78.99 & 14.94 & 5.05 & 83.59  & 24.00  \\
& SPT-Code & 17.54 & 80.25 & 14.99 & 9.39 & 87.62 & 21.51 \\
& CodeT5 & 20.36 & 80.94 & 14.62 & 10.03 & 87.51 & 20.70 \\
& \method & \textbf{23.41 ($\uparrow$ 15\%)} & \textbf{83.11} & \textbf{13.12} & \textbf{12.59 ($\uparrow$ 25.5\%)} & \textbf{91.54} & \textbf{17.59} \\ \bottomrule
\end{tabular}}
\label{tab:RQ1}
\end{table*}

\textbf{Results and Analyses.} The experimental results in two code-to-code editing datasets are shown in Table \ref{tab:RQ1}. The values in parentheses are relative improvements compared to the SOTA baseline.
We have three-fold findings.
\uline{(1) Our \method outperforms all baselines on both datasets.} Specifically, \method generates more correct programs than the SOTA baseline - CodeT5 by 15\% in the Small dataset and 25.5\% in the Medium dataset. Note that exact match is a strict metric and is difficult to be improved. The significant improvements show the superiority of our \method in automatic code editing.
\changeline{Moreover, \method obtains better results on the CrystalBLEU and Edit distance.} It shows that for incorrect predictions, the outputs of \method are closer to the ground truth and cost fewer developers' efforts to edit. 
We attribute these improvements to our pre-training task. Compared to previous code infilling tasks, we introduce a powerful language model to derive more meaningful edit patterns (\eg type/object changes, API updates) that probably occur in real-world code editing. 
These edit patterns help our \method learn the prior knowledge of code editing and strengthen the code editing ability in practical applications.
\uline{(2) Pre-trained models are more promising.}
Compared to models trained from scratch, pre-trained models achieve significant improvements. For example, in terms of the exact match, our \method improves the Transformer by 234\% in the Small dataset and 200.5\% in the Medium dataset, despite both models having the same architecture and a comparable number of parameters. The improvements show that pre-training tasks are necessary and effective for code editing.
\uline{(3) Sequence-to-sequence (Seq2Seq) learning is beneficial to code editing.}
We notice that Seq2Seq-based pre-trained models (\eg \method, CodeT5) obtain better results compared with encoder-only (\eg CodeBERT) and decoder-only (\eg CodeGPT) pre-trained models. This is because both understanding the code to be edited and correctly generating the edited code are important for code editing. The Seq2Seq pre-trained model is a better choice that contains a pre-trained encoder for understanding the code and a pre-trained decoder for generating the code.

\begin{tcolorbox}[size=title]
    \textbf{Answer to RQ1}: After being fine-tuned with two code-to-code editing datasets, \method achieves the best results among all baselines. In particular, \method generates 23.41\% and 12.59\% correct programs on two datasets, outperforming the SOTA baseline by 15\% and 25.5\%.
    The significant improvements show the effectiveness of our \method in automatic code-to-code editing. 
\end{tcolorbox}

\subsection{Performance in the comment\&code-to-code scenario}
\label{sec:result:RQ2}

\textbf{RQ2: How does \method perform in the comment\&code-to-code editing scenario?}

\textbf{Motivation.} We investigate whether our \method learns to edit source code under the guidance of a natural language comment.

\textbf{Setup.} We fine-tune or train baselines and the pre-trained \method with two comment\&code-to-code editing datasets (\ie Tufano et al. \cite{tufan2021towards} and new\_large). \changeline{Then, we use three metrics (\ie exact match, CrystalBLEU, and Edit distance) to evaluate the performance of baselines and our \method on the test data.}
Since some baselines (\ie Tufano et al. \cite{tufano2019learning}, CODIT, and AutoTansform) are designed for code-to-code editing, we omit them in this experiment.

\begin{table*}[t]
\caption{\changeline{The performance of baselines and our \method on two comment\&code-to-code editing datasets.} The values in parentheses are relative improvements compared to the SOTA baseline.}
\changeref{}
\resizebox{\linewidth}{!}{
\begin{tabular}{cl|ccc|ccc}
\toprule
\multirow{2}{*}{Type} &
  \multirow{2}{*}{Approaches} &
  \multicolumn{3}{c|}{Tufano et al. \cite{tufan2021towards} dataset} &
  \multicolumn{3}{|c}{new\_large dataset} \\ 
  & & Exact match $\uparrow$ & CrystalBLEU $\uparrow$ & Edit distance $\downarrow$
  & Exact match $\uparrow$ & CrystalBLEU $\uparrow$ & Edit distance $\downarrow$  \\ \midrule
\multirow{2}{*}{\begin{tabular}[c]{@{}c@{}}Trained from\\ Scratch\end{tabular}} 
& Tufano et al. \cite{tufan2021towards} 
& 2.93 & 49.57 & 36.70 & 0.89 & 47.87 & 63.44 \\
& Transformer 
& 7.63 & 52.13 & 33.26 & 1.31 & 51.90 & 59.06 \\
\midrule
\multirow{8}{*}{\begin{tabular}[c]{@{}c@{}}Pre-trained \\ Model\end{tabular}}
& RoBERTa (code) & 13.44 & 77.24 & 20.83 & 4.60 & 76.11 & 46.88 \\
& CodeBERT & 13.79 & 79.69 & 19.11 & 4.89 & 77.50 & 45.83 \\
& GraphCodeBERT & 21.36 & 81.29 & 17.69 & 7.73 & 80.54 & 42.97 \\
& CodeGPT & 14.96 & 82.13 & 18.77 & 4.96 & 80.05 & 44.36 \\
& T5-review & 15.26 & 83.42 & 17.28 & 5.02 & 81.73 & 42.64 \\
& SPT-Code & 23.30 & 84.79 & 14.32 & 7.78 & 82.63 & 41.39 \\
& CodeT5 & 26.88 & 85.42 & 13.75 & 7.81 & 83.32 & 39.02 \\
& \method & \textbf{29.40 ($\uparrow$ 9.4\%)} & \textbf{87.40} & \textbf{11.12} & \textbf{9.89 ($\uparrow$ 26.6\%)} & \textbf{86.46} & \textbf{36.74} \\ \bottomrule
\end{tabular}}
\label{tab:RQ2}
\end{table*}

\textbf{Results and Analyses.}
The experimental results are shown in Table \ref{tab:RQ2}. The values in parentheses are relative improvements compared to the SOTA baseline. We obtain two-fold insights. 
\uline{(1) Our \method obtains the best results among all baselines.}
Specifically, in terms of exact match, our \method outperforms the SOTA baseline - CodeT5 by 9.4\% in the Tufano et al. \cite{tufan2021towards} and 26.6\% in the new\_large dataset. We attribute these improvements to our pre-training data that contains edit pairs with comments. These edit pairs ask our model to edit the input code into a new version that is consistent with a given comment. Thus, our model learns a conditional code editing ability and performs well in comment\&code-to-code editing. 
\uline{(2) Understanding natural language comments is crucial to the comment\&code-to-code editing.}
Compared to code-to-code editing, comment\&code-to-code editing is a more challenging task that requires understanding the semantics of given natural language comments. Table \ref{tab:RQ2} shows that models trained from scratch have poor results and pre-trained models perform well. The reason is that models trained from scratch lack prior knowledge about the source code and natural languages. 
Thus, it is hard for models trained from scratch to understand the semantics of comments through the limited data. 
While pre-trained models are pre-trained with extensive code files containing natural language comments. They can learn related knowledge about the source code and natural languages, thus obtaining significant improvements.

\begin{tcolorbox}[size=title]
    \textbf{Answer to RQ2}: After being fine-tuned with two comment\&code-to-code editing datasets, \method generates 29.4\% and 9.89\% correct programs, improving the SOTA baseline by 9.4\% and 26.6\%. It shows the \method learns a high-level natural language understanding ability and a strong conditional code editing ability.
\end{tcolorbox}

\subsection{Performance in few-shot and zero-shot settings}
\label{sec:result:RQ3}

\textbf{RQ3: How does \method perform in few-shot and zero-shot settings?}

\textbf{Motivation.} We think our pre-training task is closer to real-world code editing and strengthens the model's generalization ability. Thus, we evaluate the performance of our \method in few-shot and zero-shot settings where the fine-tuning data is limited.

\textbf{Setup.} 
For the few-shot setting, we sub-sampling the code editing datasets. We fine-tune the pre-trained \method with these limited datasets and measure its performance.
For the zero-shot setting, we evaluate the pre-trained \method without fine-tuning. 
We conduct zero-shot and few-shot experiments on four datasets used in RQ1 and RQ2. To prove the superiority of our pre-training technique, we compare our \method to the SOTA pre-trained baseline - CodeT5 in terms of exact match.

Specifically, we select $r$ (\%) samples from a code editing dataset for fine-tuning and then evaluate fine-tuned models on the entire test data. 
When $r=0$, the model is not fine-tuned and evaluated on the entire test data (\ie zero-shot learning). When $r>0$, the model is fine-tuned with only a few samples and then tested on the entire test data. 
In our experiments, $r$ is set to 0\%, 10\%, 20\%, 30\%, 40\%, 50\%.

\begin{figure*}[t]
\centering
\includegraphics[width=\linewidth]{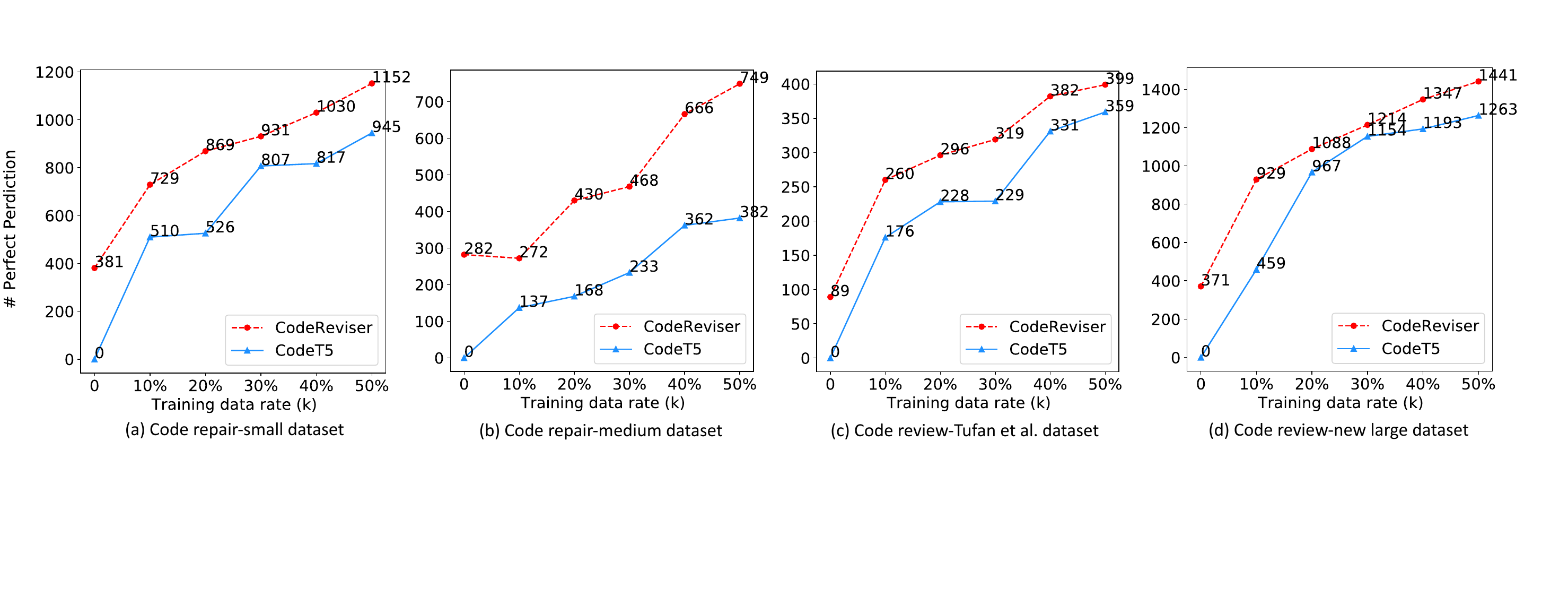}
\caption{A comparison of CodeT5 and \method in zero-shot and few-shot learning.}
\label{fig:zero_few_shot}
\end{figure*}

\begin{figure}[t]
\centering
\includegraphics[width=\linewidth]{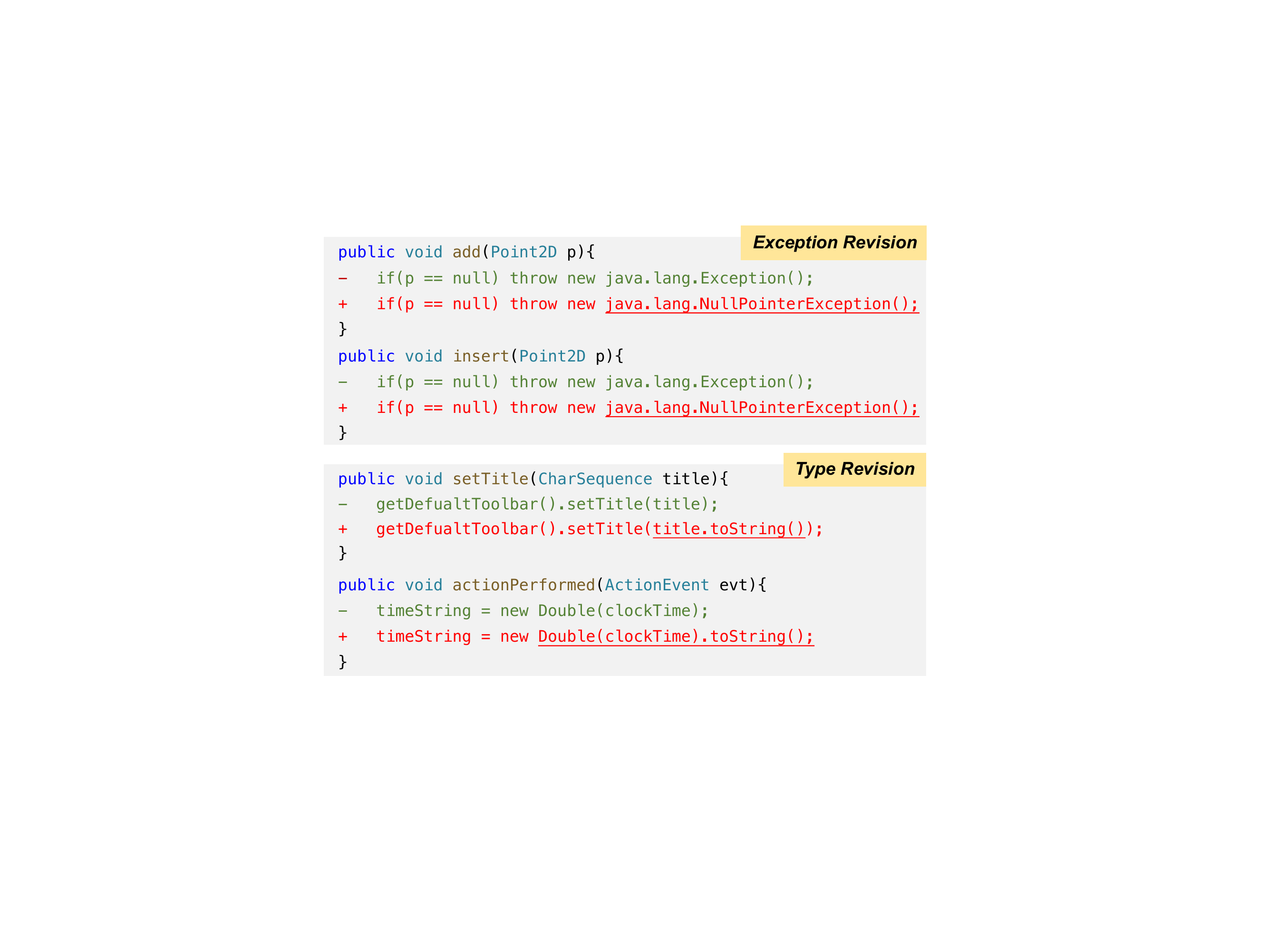}
\caption{Similar edits in our pre-training data and real-world code editing data.}
\label{fig:zero_example}
\end{figure}

\textbf{Results and Analyses.} The experimental results on four datasets are shown in Figure \ref{fig:zero_few_shot}. We have two insights. 
\uline{(1) In the zero-shot setting, our pre-trained \method successfully edits some real-world code snippets while CodeT5 can not work.} 
Specifically, in the zero-shot setting ($r=0$), our pre-trained \method correctly generates 6.53\%, 4.16\%, 5.22\%, and 2.21\% programs, for a total of 1,113 programs on four datasets. While none of the generated programs from CodeT5 is correct.
This observation validates that existing code infilling-based pre-trained models have limitations in code editing.
In this paper, we propose a novel pre-training task that is closer to real-world code editing. 
Figure \ref{fig:zero_example} shows our pre-training example and a real-world example from the Small dataset. We can see that both samples share an edit pattern. Thus, the edit patterns learned from our pre-training task can be seamlessly applied to practical applications without fine-tuning.
\uline{(2) In few-shot learning, our pre-trained \method substantially outperforms the CodeT5.}
As the data size increases, the performance of \method grows steadily and substantially outperforms CodeT5 on four datasets.
We also notice that \method fine-tuned with a few samples even outperforms several baselines fine-tuned with all samples.
For example, when being fine-tuned with 50\% samples, \method outperforms baselines that fine-tuned with all samples (\ie \method: 19.76\% \vs SPT-Code: 17.54\% on the Small dataset, \method: 11.46\% \vs CodeT5: 10.03\% on the Medium dataset, \method: 23.42\% \vs SPT-Code: 23.30\% on the Tufano et al. \cite{tufan2021towards} dataset, \method: 8.59\% \vs CodeT5: 7.81\% on the new\_large dataset). 
The results show that compared to previous pre-trained models, our pre-trained \method shows a strong generalization ability in code editing.

\begin{tcolorbox}[size=title]
    \textbf{Answer to RQ3}: In the zero-shot setting (\ie without fine-tuning), \method correctly edits 1,113 programs while CodeT5 can not work. In the few-shot setting (\ie being fine-tuned with a few samples), \method substantially outperforms CodeT5 and some baselines fine-tuned with all samples.
    It shows our pre-training task is closer to real-world code editing and strengthens the model's generalization ability.
\end{tcolorbox}

\section{Discussion}
\label{sec:discussion}

\begin{change}
\subsection{Case study}
\label{sec:discussion:qualitative}

\begin{figure}[t]
\centering
\includegraphics[width=0.8\linewidth]{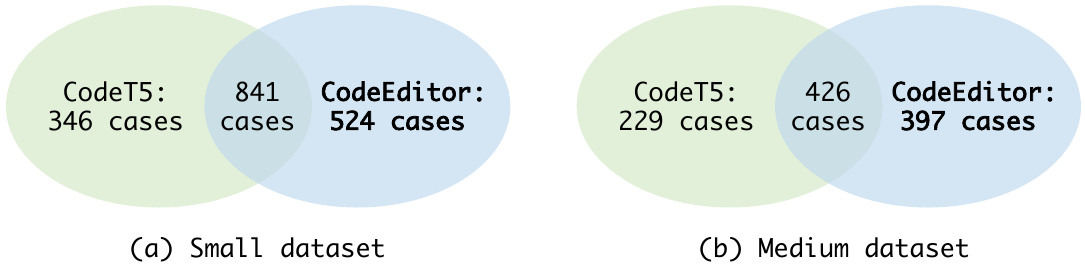}
\caption{\changeline{The statistics of successful cases on the Small and Medium datasets.}}
\label{fig:case_statistic}
\end{figure}

\paragraph{Case statistics}
We collect successful cases of CodeT5 and our \method on the Small and Medium datasets. Figure \ref{fig:case_statistic} shows the statistics of successful cases. We can see that CodeT5 and \method both can correctly solve some samples. Compared to CodeT5, \method solves unique 524 samples and 397 samples on two datasets, respectively. This result shows that \method is complementary to existing SOTA models for code editing.

\begin{figure}[t]
\centering
\includegraphics[width=\linewidth]{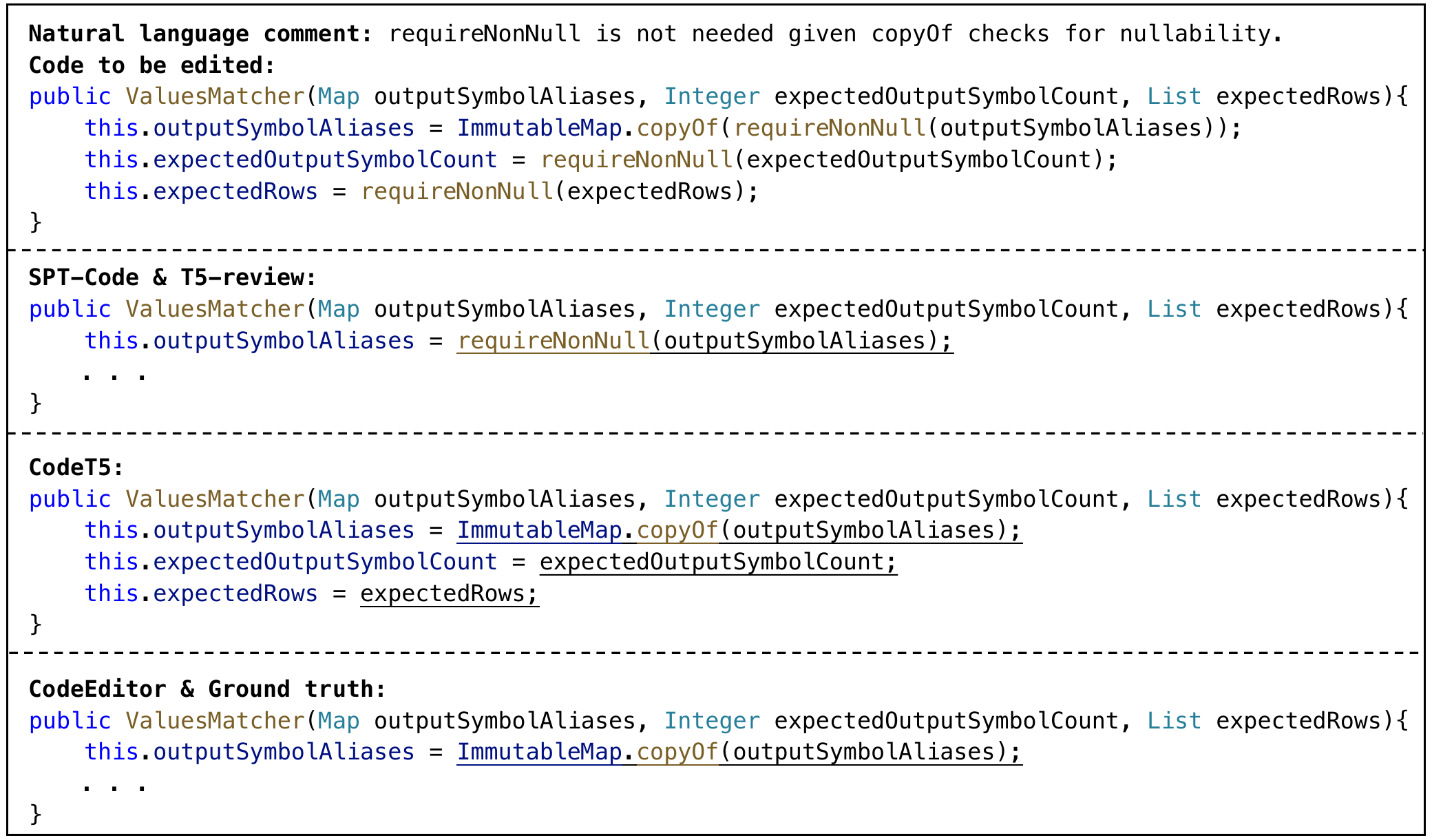}
\caption{\changeline{A successful example on the Tufano et al. \cite{tufan2021towards} dataset. To make it clear, we underline edited contents and omit some unchanged statements.}}
\label{fig:case_study}
\end{figure}

\paragraph{Successful case}
Figure \ref{fig:case_study} presents a successful example of our \method on the Tufano et al. \cite{tufan2021towards} dataset.
In this example, the natural language comment suggests removing an unnecessary call (\ie \texttt{requireNonNull}) in line 2. 
However, SPT-Code and T5-review wrongly delete the API \texttt{ImmutableMap.copyOf}, and CodeT5 mistakenly removes all \texttt{requireNonNull} calls.
This is because previous pre-trained models are pre-trained to infill a given blank instead of automatically determining edit locations based on comments. Thus, baselines often misunderstand the comment and predict wrong edits. 
While our \method is pre-trained to understand the semantics comments and edit the code with the guidance of comments.
Therefore, \method can accurately locate the line to be edited and conduct a correct edit, without introducing other defects.

\begin{figure}[t]
\centering
\includegraphics[width=0.9\linewidth]{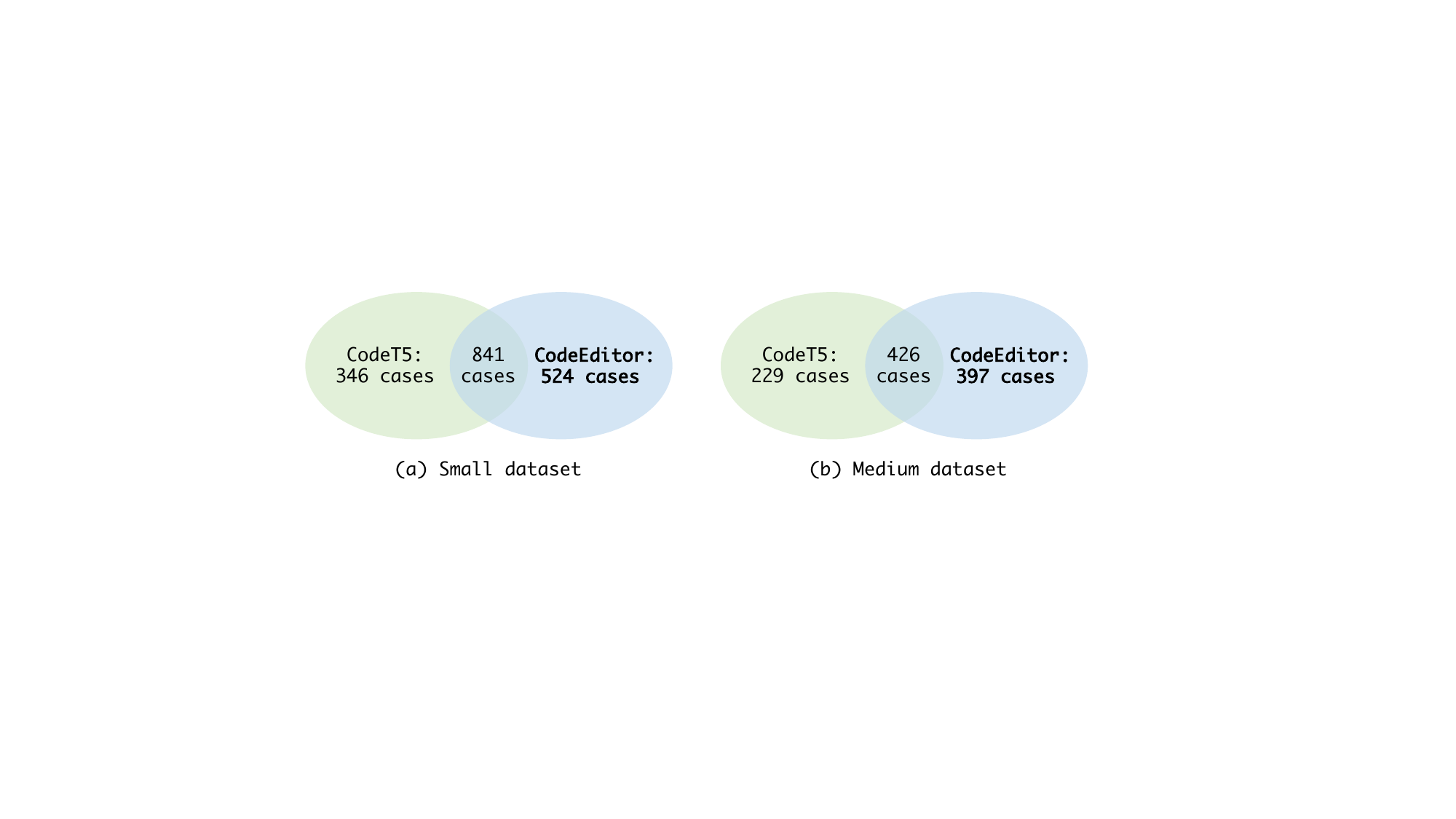}
\caption{\changeline{A failed example of \method on the Small dataset.}}
\label{fig:failed_case}
\end{figure}

\paragraph{Failed case}
Figure \ref{fig:failed_case} shows a failed example of \method on the Small dataset. In this example, the expected edit is to remove a redundant statement. While our \method mistakenly modifies this statement by adding an argument (\ie \texttt{true}). We speculate that there are two reasons for the failed example. \ding{182} \method does not how to edit. The edit in this example is rare in the data. Without additional guidance (\eg a comment), \method has no idea how to edit. It also shows that a natural language comment is important to perform code editing. \ding{183} \method tends to modify statements instead of deleting statements. As stated in Section \ref{sec:model:pre-training}, we randomly rewrite some spans of an original program to build the pre-training data. The lengths of spans are in a range of [1, 6], and the selected spans mainly are parts of statements. Thus, our pre-training data mainly is to modify statements and contains a small number of deleting complete statements. Therefore, the pre-trained \method tends to modify statements and is relatively weak in deleting complete statements. To address this problem, we will observe real-world code edits (\eg the sizes and types) and construct more effective pre-training data in future work.

\begin{figure}[t]
\centering
\includegraphics[width=\linewidth]{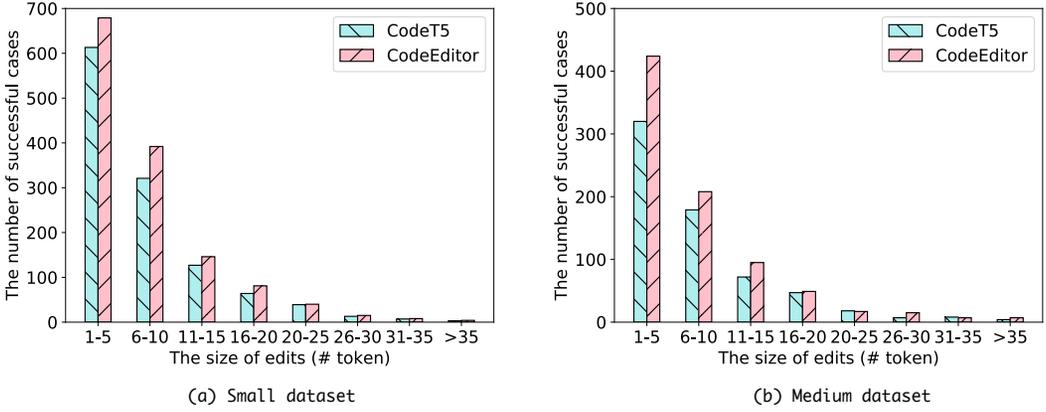}
\caption{\changeline{The performance on different edit sizes.}}
\label{fig:edit_size}
\end{figure}

\subsection{Performance on different edit sizes}
\label{sec:discussion:edit_size}

We also analyze the performance of different approaches on different (token-level) edit sizes. Specifically, we divide test data into multiple groups based on the edit sizes, such as one-token edits and two-token edits. For each group, we collect the outputs of different models and report the number of successful cases. The experimental results are shown in Figure \ref{fig:edit_size}. We can see that our \method outperforms the SOTA baseline - CodeT5 on different edit sizes. In particular, on edits of less than 10 tokens, \method keeps stable and prominent improvements over CodeT5. The results show the superiority and robustness of our model. As the edit size increase, the number of test samples decreases and the improvements of our \method are relatively slight. 

\end{change}

\subsection{Influence of more training steps}
\label{sec:discussion:steps}

Following previous studies \cite{guo2020graphcodebert,lu2021codexglue,li2022coderetriever}, we initialize our model with the weights of CodeT5, and continually pre-train the model with our pre-training task. Therefore, there is a question of whether more pre-training steps give our model an unfair advantage. To address this problem, we also continually pre-train CodeT5 using its original pre-training tasks and the number of training steps is equal to that of our model. We mark the continually pre-trained CodeT5 as CodeT5-cont.

The comparison of CodeT5-cont and our \method is shown in Table \ref{tab:steps}. Continuing pre-training brings CodeT5-cont negligible improvements and even degrades the performance on the Tufano et al. \cite{tufan2021towards} dataset. While our \method outperforms CodeT5-cont by a substantial margin on four datasets. Therefore, we can conclude that \method is still superior under the same amount of pre-training steps and data. In addition, the results demonstrate that code infilling tasks provide limited help to code editing. Our pre-training task is closer to the real-world code editing and can provide stronger supervision signals. Thus, our \method significantly improves CodeT5-cont.

\begin{table}[t]
\caption{The comparison (exact match) of CodeT5-cont and our \method.}
\begin{tabular}{l|cccc}
\toprule
  \multirow{2}{*}{Approaches} &
  \multicolumn{2}{c}{Code-to-code editing} &
  \multicolumn{2}{c}{Comment\&code-to-code editing} \\ 
  & Small & Medium & Tufano et al. \cite{tufan2021towards}
  & new\_large \\ \midrule
CodeT5 & 20.36 & 10.03 & 26.88 & 7.81 \\
CodeT5-cont & 20.40 & 10.08 & 25.53 & 8.32 \\
\method & \textbf{23.41} & \textbf{12.59} & \textbf{29.40} & \textbf{9.89} \\
\bottomrule
\end{tabular}
\label{tab:steps}
\end{table}

\subsection{Threats to validity}
\label{sec:discussion:threats}

\textbf{Threats to external validity} relate to the quality of experimental datasets and the generalizability of our results.
First, to ensure fairness of the comparison, we follow previous studies \cite{feng2020codebert,guo2020graphcodebert,wang2021codet5} and use CodeSearchNet for pre-training. The four code editing datasets for fine-tuning are mainstream benchmarks and have been used in many related works \cite{tufano2019learning,tufan2021towards,chakraborty2021MODIT,tufano2022using}. For rigorous consideration, we remove all duplicate samples between pre-training data and test data to avoid data leakage. Second, we only conduct experiments on the Java datasets. Although Java may not be representative of all programming languages, we conduct experiments on four datasets that are large and safe enough to show the effectiveness of our model. Besides, our model uses only language-agnostic features and can be applied to other programming languages.

\textbf{Threats to internal validity} include the influence of hyper-parameters. It is widely known that deep learning models are sensitive to hyper-parameters.
For the baselines, we use the source code provided by their original papers and ensure that the model's performance is comparable with their reported results.
For our \method, due to the high training cost of pre-training, we do a small-range grid search on the learning rate and batch size, leaving other hyper-parameters the same as those in previous studies \cite{wang2021codet5}. Previous work \cite{wang2021codet5} has explored effective settings of hyper-parameters through extensive experiments.
Thus, there may be room to tune more hyper-parameters for more improvements.

\textbf{Threats to construct validity} relate to the reliability of our evaluation metrics. To address this threat, we use the \textit{exact match}, \textit{CrystalBLEU} score and \textit{Edit distance} as evaluation metrics. The exact match evaluates the percentage of correctly predicted code snippets. It is a mainstream metric for code editing and is used in almost all previous studies \cite{tufano2019learning,tufan2021towards,chakraborty2021MODIT,tufano2022using, PoisonAttack}. 
Considering exact match may be too strict, we further employ CrystalBLEU and Edit distance to measure the text-similarity between predictions and the ground truth. 
Based on the above metrics, each experiment is run three times and the average result is reported.

\section{Related work}
\label{sec:related_work}

In this paper, we propose a new pre-training model for automatic code editing.
Thus, our work mainly relates to two research areas: (i) pre-training for source code, and (ii) code editing.
In this section, we summarize related work in these two areas.

\subsection{Pre-training for Source Code}
\label{sec:related_work:pre-train}

Recently, some pre-trained models for source code have been proposed and applied to a variety of software engineering (SE) tasks, such as code search \cite{feng2020codebert,guo2020graphcodebert}, code generation \cite{SkCoder,AceCoder,TiP}, and code summarization \cite{niu2022spt-code,wang2021codet5}.
Previous studies first pre-train a model with pre-training tasks, then fine-tune the pre-trained model with specific tasks.
The prior knowledge learned from pre-training tasks can enhance the performance and generalization ability of models in downstream tasks.
According to the network architecture, existing pre-trained models for source code can be categorized into three groups: encoder-only, decoder-only, and Seq2Seq-based models.

\textbf{Encoder-only} pre-trained models only contain an encoder and are trained to learn a generic code representation for code understanding tasks (\eg code classification).
CodeBERT \cite{feng2020codebert} is a pioneer encoder-only model that reuses previous pre-training tasks in the NLP filed \cite{liu2019roberta}. The obvious improvements by CodeBERT prove the effectiveness of pre-training techniques.
Guo et al. \cite{guo2020graphcodebert} further introduced data flow graphs into pre-trained models to help models understand the inherent structure of source code.

\textbf{Decoder-only} pre-trained models consist of a decoder and are mainly used for code generation tasks (\eg code completion).
Liu et al. \cite{liu2020multi} proposed a multi-task-based pre-trained model and applied it to code completion.
Inspired by the progress made by GPT models in the NLP field, some researchers \cite{lu2021codexglue} transferred the GPT-2 \cite{radford2019gpt-2} to source code and proposed a CodeGPT model, which achieved SOTA results on the code completion task.

\textbf{Seq2Seq-based} pre-trained models consist of an encoder and a decoder and can support code understanding and generation tasks.
Early studies mainly followed ideas (\eg T5 \cite{raffel2020t5}, BART \cite{lewis2020bart}) for natural languages and applied them into source code, such as T5-learning \cite{mastropaolo2021studying}, PLBART \cite{ahmad2021plbart}.
Recently, Niu et al. \cite{niu2022spt-code} designed three pre-training tasks for source code and proposed a novel pre-trained model named SPT-Code. Wang et al. \cite{wang2021codet5} proposed a pre-trained model named CodeT5 that better leverages the code semantics conveyed from the developer-assigned identiﬁers. Nowadays, CodeT5 has become the SOTA model for various code-related tasks.

\textbf{Differences.}
Although the above pre-trained models can be applied to code editing, their pre-training tasks are not designed for code editing and can be further improved.
This paper aims to propose a novel pre-training task for code editing. Our pre-training task trains a model to edit an auto-rewritten mutated code snippet into the ground truth, to learn edit patterns.
Experimental results show that our \method outperforms existing code editing baselines in three settings (\ie fine-tuning, few-shot, and zero-shot).

\begin{change}
We notice that some similar studies (\ie ELECTRA \cite{clark2019electra} and SSR \cite{zhou2021improving}) use a generator to rewrite original natural language sentences into a new version and construct the pre-training data. We think the differences between our work and these approaches are three-fold. 
\ding{182} ELECTRA \cite{clark2019electra} only rewrites single tokens in original samples and trains a model to distinguish which tokens are modified. While our \method rewrites several spans of different lengths in original samples and trains a model to reconstruct rewritten spans. Thus, compared to ELECTRA, \method can learn more edit patterns and is closer to code editing. 
\ding{183} SSR \cite{zhou2021improving} randomly rewrites several spans of original samples and uses specific tokens (\texttt{<s>}, \texttt{</s>}) to mark the locations of rewritten spans. For example, a pre-training input of SSR is \texttt{In 2001, Elon Musk joined SpaceX, <s> a manufacturer </s> company.} Then, SSR is trained to edit rewritten spans and output original spans (\eg \texttt{an aerospace manufacturer}). While our \method does not specify locations of rewritten spans. Our \method is trained to automatically locate parts that need to be edited and outputs patches. Our motivation is that code editing does not provide locations in practice. Compared to SSR, \method is closer to practical scenarios and performs better in real-world code editing data. 
\ding{184} ELECTRA \cite{clark2019electra} and SSR \cite{zhou2021improving} may produce trivial pre-training samples, \ie the mutated sample is the same as the original sample. It teaches models few edit patterns and even misleads models to copy inputs as outputs. While our \method ensures mutated samples are different from original samples. By learning from our produced data, \method knows many meaningful and diverse edit patterns.
\end{change}

\subsection{Code Editing}
\label{sec:related_work:code_editing}

The motivation of code editing is to model edit patterns in previous code edits and apply these patterns to newly-written code snippets.
It can be framed as a sequence-to-sequence task in which the code to be edited is transformed into the edited code. Nowadays, code editing mainly contains two scenarios, including code-to-code editing and comment\&code-to-code editing.

\textbf{Code-to-code editing} aims to edit a code snippet into a new version by automating generic code changes. Tufano et al. \cite{tufano2019learning, tufan2021towards} presented an initial investigation of using deep neural networks in editing the source code. They collected code changes from pull requests and proposed a standard Seq2Seq model to learn code changes. Chakraborty et al. \cite{chakraborty2020codit} further proposed a tree-based model that learned to edit the syntax tree of source code and ensured the grammatical correctness of the edited code. 
Thongtanunam et al. \cite{thongtanunam2022autotransform} introduced the BPE algorithm and the Transformer architecture to handle new tokens (\eg rare identifiers) and too-long code snippets. Recently, some powerful pre-trained models are applied to code editing and achieved SOTA results, such as SPT-Code \cite{niu2022spt-code}, T5-review \cite{tufano2022using}, and CodeT5 \cite{wang2021codet5}.

\textbf{Comment\&code-to-code editing} aims to edit a code snippet by automating code changes that are recommended by a natural language comment. Compared to code-to-code editing, comment\&code-to-code editing requires understanding the natural language comments and is a challenging task. Tufano et al. \cite{tufan2021towards} collected edit pairs from the code review and proposed a Seq2Seq model to edit developers' programs based on reviewers' comments. Inspired by the pre-training techniques, Tufano et al. \cite{tufano2022using} applied ideas \cite{raffel2020t5} in the NLP field into the source code and proposed a pre-trained code editing model named T5-review. They also released a new dataset named new\_large containing 130,000 edit pairs with comments. 
Besides, some researchers tried to apply existing pre-trained models to code editing. Chakraborty et al. \cite{chakraborty2021MODIT} considered the commit message as the guidance and utilized a pre-trained model - PLBART to edit programs.

\section{Conclusion and Future Work}
\label{sec:conclusion}

In this paper, we propose a novel pre-training task for code editing and present an effective pre-training code editing model named {\sc CodeEditor}.
Compared to previous approaches, our pre-training task endows \method with a more effective code editing ability and a stronger generalization ability.
Specifically, we collect lots of real-world code snippets as the ground truth and use a powerful generator to rewrite them into mutated versions. Then, we pre-train \method to edit mutated versions into the ground truth, to learn edit patterns. 
We conduct a large-scale study on four code editing datasets and evaluate the pre-trained \method in three settings (\ie fine-tuning, few-shot, and zero-shot). (1) In the fine-tuning setting, we train the pre-trained \method with four datasets. \method outperforms the SOTA baseline by 15\%, 25.5\%, 9.4\%, and 26.6\% on four datasets. 
(2) In zero-shot and few-shot settings where the fine-tuning data is limited, \method still substantially outperforms the SOTA baseline. 
These improvements prove that our \method is more effective in automatic code editing, and shows a strong generalization ability.
In the future, we will explore more advanced pre-training techniques for automatic code editing. For example, pre-training a model to generate edit actions.

\begin{acks}
This research is supported by the National Natural Science Foundation of China under Grant Nos. 62192733, 62192731, 61751210, 62072007, 61832009, and 62192730.
The AI training platform supporting this work was provided by High-Flyer AI. (Hangzhou High-Flyer AI Fundamental Research Co., Ltd.)
\end{acks}

\bibliographystyle{ACM-Reference-Format}
\bibliography{sample-base}


\begin{thebibliography}{40}


\ifx \showCODEN    \undefined \def \showCODEN     #1{\unskip}     \fi
\ifx \showDOI      \undefined \def \showDOI       #1{#1}\fi
\ifx \showISBNx    \undefined \def \showISBNx     #1{\unskip}     \fi
\ifx \showISBNxiii \undefined \def \showISBNxiii  #1{\unskip}     \fi
\ifx \showISSN     \undefined \def \showISSN      #1{\unskip}     \fi
\ifx \showLCCN     \undefined \def \showLCCN      #1{\unskip}     \fi
\ifx \shownote     \undefined \def \shownote      #1{#1}          \fi
\ifx \showarticletitle \undefined \def \showarticletitle #1{#1}   \fi
\ifx \showURL      \undefined \def \showURL       {\relax}        \fi
\providecommand\bibfield[2]{#2}
\providecommand\bibinfo[2]{#2}
\providecommand\natexlab[1]{#1}
\providecommand\showeprint[2][]{arXiv:#2}

\bibitem[Ahmad et~al\mbox{.}(2021)]%
        {ahmad2021plbart}
\bibfield{author}{\bibinfo{person}{Wasi Ahmad}, \bibinfo{person}{Saikat
  Chakraborty}, \bibinfo{person}{Baishakhi Ray}, {and} \bibinfo{person}{Kai-Wei
  Chang}.} \bibinfo{year}{2021}\natexlab{}.
\newblock \showarticletitle{Unified Pre-training for Program Understanding and
  Generation}. In \bibinfo{booktitle}{\emph{Proceedings of the 2021 Conference
  of the North American Chapter of the Association for Computational
  Linguistics: Human Language Technologies}}. \bibinfo{pages}{2655--2668}.
\newblock


\bibitem[Bosu and Carver(2013)]%
        {bosu2013impact}
\bibfield{author}{\bibinfo{person}{Amiangshu Bosu} {and}
  \bibinfo{person}{Jeffrey~C Carver}.} \bibinfo{year}{2013}\natexlab{}.
\newblock \showarticletitle{Impact of peer code review on peer impression
  formation: A survey}. In \bibinfo{booktitle}{\emph{2013 ACM/IEEE
  International Symposium on Empirical Software Engineering and Measurement}}.
  IEEE, \bibinfo{pages}{133--142}.
\newblock


\bibitem[Chakraborty et~al\mbox{.}(2020)]%
        {chakraborty2020codit}
\bibfield{author}{\bibinfo{person}{Saikat Chakraborty},
  \bibinfo{person}{Yangruibo Ding}, \bibinfo{person}{Miltiadis Allamanis},
  {and} \bibinfo{person}{Baishakhi Ray}.} \bibinfo{year}{2020}\natexlab{}.
\newblock \showarticletitle{Codit: Code editing with tree-based neural models}.
\newblock \bibinfo{journal}{\emph{IEEE Transactions on Software Engineering}}
  \bibinfo{volume}{48}, \bibinfo{number}{4} (\bibinfo{year}{2020}),
  \bibinfo{pages}{1385--1399}.
\newblock


\bibitem[Chakraborty and Ray(2021)]%
        {chakraborty2021MODIT}
\bibfield{author}{\bibinfo{person}{Saikat Chakraborty} {and}
  \bibinfo{person}{Baishakhi Ray}.} \bibinfo{year}{2021}\natexlab{}.
\newblock \showarticletitle{On Multi-Modal Learning of Editing Source Code}. In
  \bibinfo{booktitle}{\emph{2021 36th IEEE/ACM International Conference on
  Automated Software Engineering (ASE)}}. IEEE, \bibinfo{pages}{443--455}.
\newblock


\bibitem[Clark et~al\mbox{.}(2019)]%
        {clark2019electra}
\bibfield{author}{\bibinfo{person}{Kevin Clark}, \bibinfo{person}{Minh-Thang
  Luong}, \bibinfo{person}{Quoc~V Le}, {and} \bibinfo{person}{Christopher~D
  Manning}.} \bibinfo{year}{2019}\natexlab{}.
\newblock \showarticletitle{ELECTRA: Pre-training Text Encoders as
  Discriminators Rather Than Generators}. In
  \bibinfo{booktitle}{\emph{International Conference on Learning
  Representations}}.
\newblock


\bibitem[Devlin et~al\mbox{.}(2019)]%
        {devlin2019bert}
\bibfield{author}{\bibinfo{person}{Jacob Devlin}, \bibinfo{person}{Ming-Wei
  Chang}, \bibinfo{person}{Kenton Lee}, {and} \bibinfo{person}{Kristina
  Toutanova}.} \bibinfo{year}{2019}\natexlab{}.
\newblock \showarticletitle{BERT: Pre-training of Deep Bidirectional
  Transformers for Language Understanding}. In
  \bibinfo{booktitle}{\emph{Proceedings of the 2019 Conference of the North
  American Chapter of the Association for Computational Linguistics: Human
  Language Technologies, Volume 1 (Long and Short Papers)}}.
  \bibinfo{pages}{4171--4186}.
\newblock


\bibitem[Eghbali and Pradel(2022)]%
        {eghbali2022crystalbleu}
\bibfield{author}{\bibinfo{person}{Aryaz Eghbali} {and}
  \bibinfo{person}{Michael Pradel}.} \bibinfo{year}{2022}\natexlab{}.
\newblock \showarticletitle{CrystalBLEU: precisely and efficiently measuring
  the similarity of code}. In \bibinfo{booktitle}{\emph{37th IEEE/ACM
  International Conference on Automated Software Engineering}}.
  \bibinfo{pages}{1--12}.
\newblock


\bibitem[Feng et~al\mbox{.}(2020)]%
        {feng2020codebert}
\bibfield{author}{\bibinfo{person}{Zhangyin Feng}, \bibinfo{person}{Daya Guo},
  \bibinfo{person}{Duyu Tang}, \bibinfo{person}{Nan Duan},
  \bibinfo{person}{Xiaocheng Feng}, \bibinfo{person}{Ming Gong},
  \bibinfo{person}{Linjun Shou}, \bibinfo{person}{Bing Qin},
  \bibinfo{person}{Ting Liu}, \bibinfo{person}{Daxin Jiang}, {et~al\mbox{.}}}
  \bibinfo{year}{2020}\natexlab{}.
\newblock \showarticletitle{CodeBERT: A Pre-Trained Model for Programming and
  Natural Languages}. In \bibinfo{booktitle}{\emph{Findings of the Association
  for Computational Linguistics: EMNLP 2020}}. \bibinfo{pages}{1536--1547}.
\newblock


\bibitem[GitHub(2022)]%
        {CodeChange}
\bibfield{author}{\bibinfo{person}{GitHub}.} \bibinfo{year}{2022}\natexlab{}.
\newblock \bibinfo{title}{Real-world code changes}.
\newblock
  \bibinfo{howpublished}{\url{https://github.com/apache/hadoop/pull/4670/files\#diff-dac9de4dd225110eff2f29a44000bf32705f02df2b3fcf17b5d89bc236c12f01}}.
\newblock


\bibitem[Guo et~al\mbox{.}(2020)]%
        {guo2020graphcodebert}
\bibfield{author}{\bibinfo{person}{Daya Guo}, \bibinfo{person}{Shuo Ren},
  \bibinfo{person}{Shuai Lu}, \bibinfo{person}{Zhangyin Feng},
  \bibinfo{person}{Duyu Tang}, \bibinfo{person}{LIU Shujie},
  \bibinfo{person}{Long Zhou}, \bibinfo{person}{Nan Duan},
  \bibinfo{person}{Alexey Svyatkovskiy}, \bibinfo{person}{Shengyu Fu},
  {et~al\mbox{.}}} \bibinfo{year}{2020}\natexlab{}.
\newblock \showarticletitle{GraphCodeBERT: Pre-training Code Representations
  with Data Flow}. In \bibinfo{booktitle}{\emph{International Conference on
  Learning Representations}}.
\newblock


\bibitem[Husain et~al\mbox{.}(2019)]%
        {husain2019codesearchnet}
\bibfield{author}{\bibinfo{person}{Hamel Husain}, \bibinfo{person}{Ho{-}Hsiang
  Wu}, \bibinfo{person}{Tiferet Gazit}, \bibinfo{person}{Miltiadis Allamanis},
  {and} \bibinfo{person}{Marc Brockschmidt}.} \bibinfo{year}{2019}\natexlab{}.
\newblock \showarticletitle{CodeSearchNet Challenge: Evaluating the State of
  Semantic Code Search}.
\newblock \bibinfo{journal}{\emph{CoRR}}  \bibinfo{volume}{abs/1909.09436}
  (\bibinfo{year}{2019}).
\newblock
\showeprint[arXiv]{1909.09436}
\urldef\tempurl%
\url{http://arxiv.org/abs/1909.09436}
\showURL{%
\tempurl}


\bibitem[Lewis et~al\mbox{.}(2020)]%
        {lewis2020bart}
\bibfield{author}{\bibinfo{person}{Mike Lewis}, \bibinfo{person}{Yinhan Liu},
  \bibinfo{person}{Naman Goyal}, \bibinfo{person}{Marjan Ghazvininejad},
  \bibinfo{person}{Abdelrahman Mohamed}, \bibinfo{person}{Omer Levy},
  \bibinfo{person}{Veselin Stoyanov}, {and} \bibinfo{person}{Luke
  Zettlemoyer}.} \bibinfo{year}{2020}\natexlab{}.
\newblock \showarticletitle{BART: Denoising Sequence-to-Sequence Pre-training
  for Natural Language Generation, Translation, and Comprehension}. In
  \bibinfo{booktitle}{\emph{Proceedings of the 58th Annual Meeting of the
  Association for Computational Linguistics}}. \bibinfo{pages}{7871--7880}.
\newblock


\bibitem[Li et~al\mbox{.}(2023a)]%
        {TiP}
\bibfield{author}{\bibinfo{person}{Jia Li}, \bibinfo{person}{Ge Li},
  \bibinfo{person}{Yongmin Li}, {and} \bibinfo{person}{Zhi Jin}.}
  \bibinfo{year}{2023}\natexlab{a}.
\newblock \showarticletitle{Enabling Programming Thinking in Large Language
  Models Toward Code Generation}.
\newblock \bibinfo{journal}{\emph{CoRR}}  \bibinfo{volume}{abs/2305.06599}
  (\bibinfo{year}{2023}).
\newblock
\urldef\tempurl%
\url{https://doi.org/10.48550/arXiv.2305.06599}
\showDOI{\tempurl}
\showeprint[arXiv]{2305.06599}


\bibitem[Li et~al\mbox{.}(2021)]%
        {EditSum}
\bibfield{author}{\bibinfo{person}{Jia Li}, \bibinfo{person}{Yongmin Li},
  \bibinfo{person}{Ge Li}, \bibinfo{person}{Xing Hu}, \bibinfo{person}{Xin
  Xia}, {and} \bibinfo{person}{Zhi Jin}.} \bibinfo{year}{2021}\natexlab{}.
\newblock \showarticletitle{Editsum: A retrieve-and-edit framework for source
  code summarization}. In \bibinfo{booktitle}{\emph{2021 36th IEEE/ACM
  International Conference on Automated Software Engineering (ASE)}}. IEEE,
  \bibinfo{pages}{155--166}.
\newblock


\bibitem[Li et~al\mbox{.}(2023b)]%
        {SkCoder}
\bibfield{author}{\bibinfo{person}{Jia Li}, \bibinfo{person}{Yongmin Li},
  \bibinfo{person}{Ge Li}, \bibinfo{person}{Zhi Jin}, \bibinfo{person}{Yiyang
  Hao}, {and} \bibinfo{person}{Xing Hu}.} \bibinfo{year}{2023}\natexlab{b}.
\newblock \showarticletitle{SkCoder: {A} Sketch-based Approach for Automatic
  Code Generation}. In \bibinfo{booktitle}{\emph{45th {IEEE/ACM} International
  Conference on Software Engineering, {ICSE} 2023, Melbourne, Australia, May
  14-20, 2023}}. \bibinfo{publisher}{{IEEE}}, \bibinfo{pages}{2124--2135}.
\newblock
\urldef\tempurl%
\url{https://doi.org/10.1109/ICSE48619.2023.00179}
\showDOI{\tempurl}


\bibitem[Li et~al\mbox{.}(2022b)]%
        {PoisonAttack}
\bibfield{author}{\bibinfo{person}{Jia Li}, \bibinfo{person}{Zhuo Li},
  \bibinfo{person}{Huangzhao Zhang}, \bibinfo{person}{Ge Li},
  \bibinfo{person}{Zhi Jin}, \bibinfo{person}{Xing Hu}, {and}
  \bibinfo{person}{Xin Xia}.} \bibinfo{year}{2022}\natexlab{b}.
\newblock \showarticletitle{Poison Attack and Defense on Deep Source Code
  Processing Models}.
\newblock \bibinfo{journal}{\emph{CoRR}}  \bibinfo{volume}{abs/2210.17029}
  (\bibinfo{year}{2022}).
\newblock
\urldef\tempurl%
\url{https://doi.org/10.48550/arXiv.2210.17029}
\showDOI{\tempurl}
\showeprint[arXiv]{2210.17029}


\bibitem[Li et~al\mbox{.}(2023c)]%
        {AceCoder}
\bibfield{author}{\bibinfo{person}{Jia Li}, \bibinfo{person}{Yunfei Zhao},
  \bibinfo{person}{Yongmin Li}, \bibinfo{person}{Ge Li}, {and}
  \bibinfo{person}{Zhi Jin}.} \bibinfo{year}{2023}\natexlab{c}.
\newblock \showarticletitle{Towards Enhancing In-Context Learning for Code
  Generation}.
\newblock \bibinfo{journal}{\emph{CoRR}}  \bibinfo{volume}{abs/2303.17780}
  (\bibinfo{year}{2023}).
\newblock
\urldef\tempurl%
\url{https://doi.org/10.48550/arXiv.2303.17780}
\showDOI{\tempurl}
\showeprint[arXiv]{2303.17780}


\bibitem[Li et~al\mbox{.}(2022a)]%
        {li2022coderetriever}
\bibfield{author}{\bibinfo{person}{Xiaonan Li}, \bibinfo{person}{Yeyun Gong},
  \bibinfo{person}{Yelong Shen}, \bibinfo{person}{Xipeng Qiu},
  \bibinfo{person}{Hang Zhang}, \bibinfo{person}{Bolun Yao},
  \bibinfo{person}{Weizhen Qi}, \bibinfo{person}{Daxin Jiang},
  \bibinfo{person}{Weizhu Chen}, {and} \bibinfo{person}{Nan Duan}.}
  \bibinfo{year}{2022}\natexlab{a}.
\newblock \showarticletitle{CodeRetriever: Unimodal and Bimodal Contrastive
  Learning}.
\newblock \bibinfo{journal}{\emph{CoRR}}  \bibinfo{volume}{abs/2201.10866}
  (\bibinfo{year}{2022}).
\newblock
\showeprint[arXiv]{2201.10866}
\urldef\tempurl%
\url{https://arxiv.org/abs/2201.10866}
\showURL{%
\tempurl}


\bibitem[Liu et~al\mbox{.}(2020)]%
        {liu2020multi}
\bibfield{author}{\bibinfo{person}{Fang Liu}, \bibinfo{person}{Ge Li},
  \bibinfo{person}{Yunfei Zhao}, {and} \bibinfo{person}{Zhi Jin}.}
  \bibinfo{year}{2020}\natexlab{}.
\newblock \showarticletitle{Multi-task learning based pre-trained language
  model for code completion}. In \bibinfo{booktitle}{\emph{Proceedings of the
  35th IEEE/ACM International Conference on Automated Software Engineering}}.
  \bibinfo{pages}{473--485}.
\newblock


\bibitem[Liu et~al\mbox{.}(2019)]%
        {liu2019roberta}
\bibfield{author}{\bibinfo{person}{Yinhan Liu}, \bibinfo{person}{Myle Ott},
  \bibinfo{person}{Naman Goyal}, \bibinfo{person}{Jingfei Du},
  \bibinfo{person}{Mandar Joshi}, \bibinfo{person}{Danqi Chen},
  \bibinfo{person}{Omer Levy}, \bibinfo{person}{Mike Lewis},
  \bibinfo{person}{Luke Zettlemoyer}, {and} \bibinfo{person}{Veselin
  Stoyanov}.} \bibinfo{year}{2019}\natexlab{}.
\newblock \showarticletitle{RoBERTa: {A} Robustly Optimized {BERT} Pretraining
  Approach}.
\newblock \bibinfo{journal}{\emph{CoRR}}  \bibinfo{volume}{abs/1907.11692}
  (\bibinfo{year}{2019}).
\newblock
\showeprint[arXiv]{1907.11692}
\urldef\tempurl%
\url{http://arxiv.org/abs/1907.11692}
\showURL{%
\tempurl}


\bibitem[Lu et~al\mbox{.}(2021)]%
        {lu2021codexglue}
\bibfield{author}{\bibinfo{person}{Shuai Lu}, \bibinfo{person}{Daya Guo},
  \bibinfo{person}{Shuo Ren}, \bibinfo{person}{Junjie Huang},
  \bibinfo{person}{Alexey Svyatkovskiy}, \bibinfo{person}{Ambrosio Blanco},
  \bibinfo{person}{Colin Clement}, \bibinfo{person}{Dawn Drain},
  \bibinfo{person}{Daxin Jiang}, \bibinfo{person}{Duyu Tang}, {et~al\mbox{.}}}
  \bibinfo{year}{2021}\natexlab{}.
\newblock \showarticletitle{CodeXGLUE: A Machine Learning Benchmark Dataset for
  Code Understanding and Generation}. In \bibinfo{booktitle}{\emph{Thirty-fifth
  Conference on Neural Information Processing Systems Datasets and Benchmarks
  Track (Round 1)}}.
\newblock


\bibitem[Mastropaolo et~al\mbox{.}(2021)]%
        {mastropaolo2021studying}
\bibfield{author}{\bibinfo{person}{Antonio Mastropaolo},
  \bibinfo{person}{Simone Scalabrino}, \bibinfo{person}{Nathan Cooper},
  \bibinfo{person}{David~Nader Palacio}, \bibinfo{person}{Denys Poshyvanyk},
  \bibinfo{person}{Rocco Oliveto}, {and} \bibinfo{person}{Gabriele Bavota}.}
  \bibinfo{year}{2021}\natexlab{}.
\newblock \showarticletitle{Studying the usage of text-to-text transfer
  transformer to support code-related tasks}. In \bibinfo{booktitle}{\emph{2021
  IEEE/ACM 43rd International Conference on Software Engineering (ICSE)}}.
  IEEE, \bibinfo{pages}{336--347}.
\newblock


\bibitem[Nguyen et~al\mbox{.}(2016)]%
        {nguyen2016api}
\bibfield{author}{\bibinfo{person}{Anh~Tuan Nguyen}, \bibinfo{person}{Michael
  Hilton}, \bibinfo{person}{Mihai Codoban}, \bibinfo{person}{Hoan~Anh Nguyen},
  \bibinfo{person}{Lily Mast}, \bibinfo{person}{Eli Rademacher},
  \bibinfo{person}{Tien~N Nguyen}, {and} \bibinfo{person}{Danny Dig}.}
  \bibinfo{year}{2016}\natexlab{}.
\newblock \showarticletitle{API code recommendation using statistical learning
  from fine-grained changes}. In \bibinfo{booktitle}{\emph{Proceedings of the
  2016 24th ACM SIGSOFT International Symposium on Foundations of Software
  Engineering}}. \bibinfo{pages}{511--522}.
\newblock


\bibitem[Nguyen et~al\mbox{.}(2013)]%
        {nguyen2013study}
\bibfield{author}{\bibinfo{person}{Hoan~Anh Nguyen}, \bibinfo{person}{Anh~Tuan
  Nguyen}, \bibinfo{person}{Tung~Thanh Nguyen}, \bibinfo{person}{Tien~N
  Nguyen}, {and} \bibinfo{person}{Hridesh Rajan}.}
  \bibinfo{year}{2013}\natexlab{}.
\newblock \showarticletitle{A study of repetitiveness of code changes in
  software evolution}. In \bibinfo{booktitle}{\emph{2013 28th IEEE/ACM
  International Conference on Automated Software Engineering (ASE)}}. IEEE,
  \bibinfo{pages}{180--190}.
\newblock


\bibitem[Niu et~al\mbox{.}(2022)]%
        {niu2022spt-code}
\bibfield{author}{\bibinfo{person}{Changan Niu}, \bibinfo{person}{Chuanyi Li},
  \bibinfo{person}{Vincent Ng}, \bibinfo{person}{Jidong Ge},
  \bibinfo{person}{Liguo Huang}, {and} \bibinfo{person}{Bin Luo}.}
  \bibinfo{year}{2022}\natexlab{}.
\newblock \showarticletitle{SPT-Code: Sequence-to-Sequence Pre-Training for
  Learning the Representation of Source Code}. In
  \bibinfo{booktitle}{\emph{2022 IEEE/ACM 44st International Conference on
  Software Engineering (ICSE)}}. IEEE.
\newblock


\bibitem[Papineni et~al\mbox{.}(2002)]%
        {papineni2002bleu}
\bibfield{author}{\bibinfo{person}{Kishore Papineni}, \bibinfo{person}{Salim
  Roukos}, \bibinfo{person}{Todd Ward}, {and} \bibinfo{person}{Wei-Jing Zhu}.}
  \bibinfo{year}{2002}\natexlab{}.
\newblock \showarticletitle{Bleu: a method for automatic evaluation of machine
  translation}. In \bibinfo{booktitle}{\emph{Proceedings of the 40th annual
  meeting of the Association for Computational Linguistics}}.
  \bibinfo{pages}{311--318}.
\newblock


\bibitem[Radford et~al\mbox{.}(2019)]%
        {radford2019gpt-2}
\bibfield{author}{\bibinfo{person}{Alec Radford}, \bibinfo{person}{Jeffrey Wu},
  \bibinfo{person}{Rewon Child}, \bibinfo{person}{David Luan},
  \bibinfo{person}{Dario Amodei}, {and} \bibinfo{person}{Ilya Sutskever}.}
  \bibinfo{year}{2019}\natexlab{}.
\newblock \showarticletitle{Language models are unsupervised multitask
  learners}.
\newblock \bibinfo{journal}{\emph{OpenAI blog}} \bibinfo{volume}{1},
  \bibinfo{number}{8} (\bibinfo{year}{2019}), \bibinfo{pages}{9}.
\newblock
\urldef\tempurl%
\url{https://cdn.openai.com/better-language-models/language_models_are_unsupervised_multitask_learners.pdf}
\showURL{%
\tempurl}


\bibitem[Raffel et~al\mbox{.}(2020)]%
        {raffel2020t5}
\bibfield{author}{\bibinfo{person}{Colin Raffel}, \bibinfo{person}{Noam
  Shazeer}, \bibinfo{person}{Adam Roberts}, \bibinfo{person}{Katherine Lee},
  \bibinfo{person}{Sharan Narang}, \bibinfo{person}{Michael Matena},
  \bibinfo{person}{Yanqi Zhou}, \bibinfo{person}{Wei Li}, {and}
  \bibinfo{person}{Peter~J Liu}.} \bibinfo{year}{2020}\natexlab{}.
\newblock \showarticletitle{Exploring the Limits of Transfer Learning with a
  Unified Text-to-Text Transformer}.
\newblock \bibinfo{journal}{\emph{Journal of Machine Learning Research}}
  \bibinfo{volume}{21} (\bibinfo{year}{2020}), \bibinfo{pages}{1--67}.
\newblock


\bibitem[Ray et~al\mbox{.}(2013)]%
        {ray2013detecting}
\bibfield{author}{\bibinfo{person}{Baishakhi Ray}, \bibinfo{person}{Miryung
  Kim}, \bibinfo{person}{Suzette Person}, {and} \bibinfo{person}{Neha Rungta}.}
  \bibinfo{year}{2013}\natexlab{}.
\newblock \showarticletitle{Detecting and characterizing semantic
  inconsistencies in ported code}. In \bibinfo{booktitle}{\emph{2013 28th
  IEEE/ACM International Conference on Automated Software Engineering (ASE)}}.
  IEEE, \bibinfo{pages}{367--377}.
\newblock


\bibitem[Sennrich et~al\mbox{.}(2016)]%
        {sennrich2016neural}
\bibfield{author}{\bibinfo{person}{Rico Sennrich}, \bibinfo{person}{Barry
  Haddow}, {and} \bibinfo{person}{Alexandra Birch}.}
  \bibinfo{year}{2016}\natexlab{}.
\newblock \showarticletitle{Neural Machine Translation of Rare Words with
  Subword Units}. In \bibinfo{booktitle}{\emph{Proceedings of the 54th Annual
  Meeting of the Association for Computational Linguistics (Volume 1: Long
  Papers)}}. \bibinfo{pages}{1715--1725}.
\newblock


\bibitem[Strubell et~al\mbox{.}(2019)]%
        {strubell2019energy}
\bibfield{author}{\bibinfo{person}{Emma Strubell}, \bibinfo{person}{Ananya
  Ganesh}, {and} \bibinfo{person}{Andrew McCallum}.}
  \bibinfo{year}{2019}\natexlab{}.
\newblock \showarticletitle{Energy and Policy Considerations for Deep Learning
  in NLP}. In \bibinfo{booktitle}{\emph{Proceedings of the 57th Annual Meeting
  of the Association for Computational Linguistics}}.
  \bibinfo{pages}{3645--3650}.
\newblock


\bibitem[Sun et~al\mbox{.}(2020)]%
        {sun2020treegen}
\bibfield{author}{\bibinfo{person}{Zeyu Sun}, \bibinfo{person}{Qihao Zhu},
  \bibinfo{person}{Yingfei Xiong}, \bibinfo{person}{Yican Sun},
  \bibinfo{person}{Lili Mou}, {and} \bibinfo{person}{Lu Zhang}.}
  \bibinfo{year}{2020}\natexlab{}.
\newblock \showarticletitle{Treegen: A tree-based transformer architecture for
  code generation}. In \bibinfo{booktitle}{\emph{Proceedings of the AAAI
  Conference on Artificial Intelligence}}, Vol.~\bibinfo{volume}{34}.
  \bibinfo{pages}{8984--8991}.
\newblock


\bibitem[Thongtanunam et~al\mbox{.}(2022)]%
        {thongtanunam2022autotransform}
\bibfield{author}{\bibinfo{person}{Patanamon Thongtanunam},
  \bibinfo{person}{Chanathip Pornprasit}, {and} \bibinfo{person}{Chakkrit
  Tantithamthavorn}.} \bibinfo{year}{2022}\natexlab{}.
\newblock \showarticletitle{AutoTransform: Automated Code Transformation to
  Support Modern Code Review Process}. In \bibinfo{booktitle}{\emph{2022
  IEEE/ACM 44st International Conference on Software Engineering (ICSE)}}.
  IEEE.
\newblock


\bibitem[Tufano et~al\mbox{.}(2019)]%
        {tufano2019learning}
\bibfield{author}{\bibinfo{person}{Michele Tufano}, \bibinfo{person}{Jevgenija
  Pantiuchina}, \bibinfo{person}{Cody Watson}, \bibinfo{person}{Gabriele
  Bavota}, {and} \bibinfo{person}{Denys Poshyvanyk}.}
  \bibinfo{year}{2019}\natexlab{}.
\newblock \showarticletitle{On learning meaningful code changes via neural
  machine translation}. In \bibinfo{booktitle}{\emph{2019 IEEE/ACM 41st
  International Conference on Software Engineering (ICSE)}}. IEEE,
  \bibinfo{pages}{25--36}.
\newblock


\bibitem[Tufano et~al\mbox{.}(2022)]%
        {tufano2022using}
\bibfield{author}{\bibinfo{person}{Rosalia Tufano}, \bibinfo{person}{Simone
  Masiero}, \bibinfo{person}{Antonio Mastropaolo}, \bibinfo{person}{Luca
  Pascarella}, \bibinfo{person}{Denys Poshyvanyk}, {and}
  \bibinfo{person}{Gabriele Bavota}.} \bibinfo{year}{2022}\natexlab{}.
\newblock \showarticletitle{Using pre-trained models to boost code review
  automation}. In \bibinfo{booktitle}{\emph{Proceedings of the 44th
  International Conference on Software Engineering}}.
  \bibinfo{pages}{2291--2302}.
\newblock


\bibitem[Tufano et~al\mbox{.}(2021)]%
        {tufan2021towards}
\bibfield{author}{\bibinfo{person}{Rosalia Tufano}, \bibinfo{person}{Luca
  Pascarella}, \bibinfo{person}{Michele Tufanoy}, \bibinfo{person}{Denys
  Poshyvanykz}, {and} \bibinfo{person}{Gabriele Bavota}.}
  \bibinfo{year}{2021}\natexlab{}.
\newblock \showarticletitle{Towards automating code review activities}. In
  \bibinfo{booktitle}{\emph{2021 IEEE/ACM 43rd International Conference on
  Software Engineering (ICSE)}}. IEEE, \bibinfo{pages}{163--174}.
\newblock


\bibitem[Vaswani et~al\mbox{.}(2017)]%
        {vaswani2017attention}
\bibfield{author}{\bibinfo{person}{Ashish Vaswani}, \bibinfo{person}{Noam
  Shazeer}, \bibinfo{person}{Niki Parmar}, \bibinfo{person}{Jakob Uszkoreit},
  \bibinfo{person}{Llion Jones}, \bibinfo{person}{Aidan~N. Gomez},
  \bibinfo{person}{Lukasz Kaiser}, {and} \bibinfo{person}{Illia Polosukhin}.}
  \bibinfo{year}{2017}\natexlab{}.
\newblock \showarticletitle{Attention is All you Need}. In
  \bibinfo{booktitle}{\emph{Advances in Neural Information Processing Systems
  30: Annual Conference on Neural Information Processing Systems 2017, December
  4-9, 2017, Long Beach, CA, {USA}}},
  \bibfield{editor}{\bibinfo{person}{Isabelle Guyon}, \bibinfo{person}{Ulrike
  von Luxburg}, \bibinfo{person}{Samy Bengio}, \bibinfo{person}{Hanna~M.
  Wallach}, \bibinfo{person}{Rob Fergus}, \bibinfo{person}{S.~V.~N.
  Vishwanathan}, {and} \bibinfo{person}{Roman Garnett}} (Eds.).
  \bibinfo{pages}{5998--6008}.
\newblock
\urldef\tempurl%
\url{https://proceedings.neurips.cc/paper/2017/hash/3f5ee243547dee91fbd053c1c4a845aa-Abstract.html}
\showURL{%
\tempurl}


\bibitem[Wang et~al\mbox{.}(2021)]%
        {wang2021codet5}
\bibfield{author}{\bibinfo{person}{Yue Wang}, \bibinfo{person}{Weishi Wang},
  \bibinfo{person}{Shafiq Joty}, {and} \bibinfo{person}{Steven~CH Hoi}.}
  \bibinfo{year}{2021}\natexlab{}.
\newblock \showarticletitle{CodeT5: Identifier-aware Unified Pre-trained
  Encoder-Decoder Models for Code Understanding and Generation}. In
  \bibinfo{booktitle}{\emph{Proceedings of the 2021 Conference on Empirical
  Methods in Natural Language Processing}}. \bibinfo{pages}{8696--8708}.
\newblock


\bibitem[Yin and Neubig(2018)]%
        {yin2018tranx}
\bibfield{author}{\bibinfo{person}{Pengcheng Yin} {and} \bibinfo{person}{Graham
  Neubig}.} \bibinfo{year}{2018}\natexlab{}.
\newblock \showarticletitle{TRANX: A Transition-based Neural Abstract Syntax
  Parser for Semantic Parsing and Code Generation}. In
  \bibinfo{booktitle}{\emph{Proceedings of the Conference on Empirical Methods
  in Natural Language Processing (Demo Track)}}.
\newblock


\bibitem[Zhou et~al\mbox{.}(2021)]%
        {zhou2021improving}
\bibfield{author}{\bibinfo{person}{Wangchunshu Zhou}, \bibinfo{person}{Tao Ge},
  \bibinfo{person}{Canwen Xu}, \bibinfo{person}{Ke Xu}, {and}
  \bibinfo{person}{Furu Wei}.} \bibinfo{year}{2021}\natexlab{}.
\newblock \showarticletitle{Improving Sequence-to-Sequence Pre-training via
  Sequence Span Rewriting}. In \bibinfo{booktitle}{\emph{Proceedings of the
  2021 Conference on Empirical Methods in Natural Language Processing}}.
  \bibinfo{pages}{571--582}.
\newblock


\end{thebibliography}

\end{document}